# Age and Cognitive Skills: Use It or Lose It[*]

Eric A. Hanushek, Lavinia Kinne, Frauke Witthöft, and Ludger Woessmann[†]

**Abstract**

Cross-sectional age-skill profiles suggest that workers' cognitive skills start declining by their thirties if not earlier. If accurate, such age-driven skill losses pose a major threat to the human capital of societies with rapidly aging populations. We estimate actual age-skill profiles from individual changes in skills at different ages. We use the unique German longitudinal component of the Programme of the International Assessment of Adult Competencies (PIAAC-L) that retested a large representative sample of adults after 3.5 years. Two main results emerge. First, correcting for measurement error, average skills increase into the forties before decreasing slightly in literacy and more strongly in numeracy. Second, skills decline at older ages only for those with below-average skill usage. White-collar and higher-educated workers with above-average usage show increasing skills even beyond their forties. Women have larger skill losses at older age, particularly in numeracy.

July 17, 2024

---

[*] We thank seminar participants at the University of Mannheim and the ifo Institute for helpful comments and discussion. This work was supported by the Smith Richardson Foundation. The contribution by Woessmann is also part of German Science Foundation project CRC TRR 190.

[†] Hanushek: Hoover Institution, Stanford University; CESifo, IZA, and NBER, hanushek@stanford.edu. Kinne: DIW Berlin and University of Potsdam, lkinne@diw.de. Witthöft: ifo Institute at the University of Munich, witthoeft@ifo.de. Woessmann: University of Munich, ifo Institute; Hoover Institution, Stanford University; CESifo, and IZA, woessmann@ifo.de.

# 1. Introduction

Increasing attention to workers' skills has prompted new questions about the nature and evolution of skills. The conventional assumption is that cognitive skills decline with age starting rather early in adult life (e.g., Desjardins and Warnke (2012)). The close relationship between general cognitive skills and economic outcomes[1] implies that negative age-skill patterns may foreshadow future economic changes emanating from the steady and dramatic changes in the age composition of societies (Bloom and Zucker (2023)). But this assumed skill pattern has largely come from cross-sectional data that necessarily incorporate not only aging patterns but also cohort differences in skills. If the negative age pattern is simply due to conflating age and cohort effects, the economic concerns are considerably lessened. Recently developed individual longitudinal skill tests for a representative population of adults allow us to provide new evidence on actual changes in skills with age and on their relationship with adult skill usage.

We study the prevalence and inevitability of age-driven skill decline using exceptional longitudinal data for the German adult population. The Programme for the International Assessment of Adult Competencies (PIAAC) tested skills that are relevant for participation in social life and work and surveyed economic and social conditions for random samples of the population aged 16-65 in 39 countries (OECD (2013)).[2] Germany, unique among all participating countries, created a panel of participants from the PIAAC sample who were re-surveyed *and* retested 3.5 years after the original survey.

We use the panel dimension of the skill data to estimate credible age-skill profiles for the average adult in the population. Observing actual changes in adult skills over the entire age spectrum allows us to break the confounding of age and cohort patterns that has been ubiquitous in general cross-sectional analyses of adult skills.

Creating reliable age-skill patterns from panel data on individual aging requires addressing the bias from measurement errors that inevitably accompany testing of individual skills over time. Observations of test scores include a combination of true scores and measurement errors, leading to systematic errors when looking at individual changes with age. Intuitively, low observed test scores are more likely to include negative errors. When we observe another assessment for an initially low-scoring individual, the measurement error is unlikely to be as negative as the first time, implying that the change in test scores over time is likely to be biased upward for low-scoring individuals. For initially high-scoring individuals, the opposite will be true. This reversion to the mean will bias the overall observed age-skill relationship when skills vary by age. Thus, we correct the observed change in test scores throughout for reversion to the mean (Berry et al. (1984)) in order to obtain error-adjusted age-skill patterns.

---

[1] Hanushek and Woessmann (2008) review early analyses of the economic role of cognitive skills. More recent international analyses show the strong impacts on individual earnings (Hanushek et al. (2015, 2017), Hampf, Wiederhold, and Woessmann (2017)) and on national growth rates (Hanushek and Woessmann (2015)).

[2] This survey design extended two previous international surveys for more limited sets of countries. Between 1994 and 1998, the International Adult Literacy Survey (IALS) was the first ever large-scale comparative assessment focusing on adult competencies in language and math. This was followed by the 2003 Adult Literacy and Life Skills Study (ALL) that employed the same design for a smaller set of countries.

We relate the estimated age-skill profiles to differences in the usage of skills. Previous analyses have investigated whether individual background or jobs influence the evolution of age-skill patterns. A comprehensive analysis of background factors would naturally begin with the adult usage of skills, but data on skill usage have not been generally available. Importantly, the background data from the PIAAC survey provide information on the detailed nature and frequency of participants' skill usage at work and at home.

We find that average (error-corrected) skills increase substantially into the forties for both literacy and numeracy. Subsequently, average skills decline slightly in literacy and strongly in numeracy. Both patterns, however, mask the fact that skills of individuals in the upper half of the population in terms of skill usage at work and home *never* decline.

The age-skill pattern is highly correlated with a variety of background characteristics, but these relationships interact closely with skill usage. For example, both literacy and numeracy skills keep increasing for white-collar and tertiary-educated workers in the second half of their working life if they have above-median skill usage, but not if skill usage is below the median.

The primary contribution of our analysis is the development of credible and generalizable age-skill profiles that relate to the literature on the economic effects of general cognitive skills. Based on these, we then also describe the connection of individual differences with behavior and background.

The next section describes the relevant related literature. Sections 3 and 4 introduce our empirical approach and the data. Sections 5 and 6 present the results on the age-skill profile and on its relationship with skill usage, occupation, education, and gender. Section 7 concludes.

## 2. Related Literature on Age-skill Analyses

The available literature comes from a variety of perspectives and designs, making it difficult to generalize from the separate studies. The studies closest to our analysis employ large-scale representative surveys whose measures of general cognitive skills have been shown to be linked to economic outcomes. A parallel set of psychological and neurological studies offer related estimates of age patterns, albeit for subsets of specific skills and for unrepresentative samples.

A series of studies has come from the prior international adult surveys (IALS and ALL). While they consider skills of nationally representative samples, they rely on purely cross-sectional information for estimating age-skill patterns, requiring the strong assumption that individuals born and educated over a long period of time are otherwise similar in the factors affecting cognitive skills (for details of the various studies, see the review by Desjardins and Warnke (2012)). An alternative has been to combine IALS, ALL, and PIAAC data to create synthetic cohorts that allow tracing skills over time for separate representative samples of birth cohorts (e.g., Green and Riddell (2013), Barrett and Riddell (2019), Reiter (2022)). However, these face problems with varying sampling errors, and lack of psychometric linkage undermines comparability of skill measures from the different tests, leading to possibly erroneous conclusions from synthetic cohort methods (Vézina and Bélanger (2020)). An exception to the cross-sectional and synthetic panel studies is a small second testing of the Swedish IALS data on a slightly different follow-on assessment (622 usable retests in 1998 of the original 3,038 observations in 1994) used by Edin and Gustavsson (2008). This study, however, focuses on skill depreciation during unemployment rather than overall age-skill profiles.

The extensive existing body of psychological and neurological research has not focused on general skills and has relied on convenience samples. It indicates that age patterns can differ markedly for different components of cognitive functioning (see Desjardins and Warnke (2012) and Sánchez-Izquierdo and Fernández-Ballesteros (2021) for reviews). While findings vary, both cross-sectional and longitudinal studies tend to show that components related to fluid intelligence (independent of prior learning), such as attentional and memory capacity, processing speed, and spatial orientation, tend to start declining very early in adult life, whereas components related to crystallized intelligence (deduced from prior learning), such as vocabulary knowledge and general information, tend to peak at later ages (e.g., Horn and Cattell (1967), Baltes (1993), Schaie (2005), Salthouse (2010b)). Given these different patterns, it is not clear what to deduce for the age trajectory of more general literacy and numeracy skills. Existing longitudinal results from these studies on cognitive aging also tend to come from unrepresentative samples. For example, the groundbreaking Seattle Longitudinal Study follows convenience samples from a Health Maintenance Organization (HMO) in Washington State since 1956 (Schaie (2005)). Limited representation may constrain generalizability, as age-skill patterns likely differ markedly with behavioral and contextual factors, as shown below. Lack of individually repeated testing for a representative adult population has been the main impediment to understanding the age patterns in skills. Our analysis contributes to this literature by studying age-skill patterns using longitudinal variation in literacy and numeracy skills in a representative adult sample.

## 3. Empirical Approach

The conventional wisdom that forms the starting point for our analysis comes from the cross-sectional picture of how general skills vary with age. Figure 1 depicts adult skills for a representative sample of the German population in our data, the PIAAC test. In the cross-section, literacy scores steadily fall on average from age 20 while numeracy scores rise slightly before beginning to decline in the late thirties. This pattern for Germany is duplicated qualitatively across the pooled sample of all OECD countries that participated in PIAAC (see Appendix Figure A1).[3]

However, it is difficult to interpret these cross-sectional patterns because they conflate age and cohort effects. Individuals across different ages also come from different cohorts and have thus experienced different histories of skill determinants. Thus, these charts fail to describe the likely patterns of age-skill relationships for any individuals. For that reason, they also cannot reliably be used to understand the factors that feed into skill changes by age without introducing strong assumptions about the nature of cohort or individual time patterns.[4]

**Changes in skills in individual longitudinal data.** Our analysis uses individual data about how literacy and numeracy scores change over a 3-to-4-year period to trace how skills change within individuals across the entire age spectrum.

---

[3] See OECD (2013) for similar depictions. A comparable cross-sectional pattern motivates, for example, the analysis in Arellano-Bover (2022) (see his Figure 1).

[4] The analysis of synthetic cohorts that comes from merging data across different samples (e.g., IALS and ALL) solves some of these problems but introduces larger sampling errors and must deal with different skill assessments.

We observe measures of skills, $T_{ia}$, for individual $i$ at age $a$ along with a second measure of the same skills at age $a^*$. We are interested in estimating the average change in skills as individuals move from $a$ to $a^*$. In the absence of measurement error, one could take the simple average of the individual differences in skills at any given age:

$$\Delta_a = T_{ia^*} - T_{ia} \tag{1}$$

These skill changes, observed within individuals, are not confounded with cohort effects.

The observed individual changes in skills with aging can emanate from different sources. They may stem from biological factors such as neurological changes as well as from behavioral factors such as physical, social, and mental activity including education, training, and skill usage (e.g., Desjardins and Warnke (2012)). These factors may also interact with environmental and social influences such as physical and cultural context. To assess how skills change with aging, we are interested in the overall skill changes emerging from the combination of these biological, behavioral, and contextual sources.

**Adjustment for measurement error due to reversion to the mean.** Analyzing the pattern of changes in skills for individuals, however, leads to immediate measurement issues. The complication is that the test-score measure of skill comes with error such that:

$$T_{ia} = \tilde{T}_{ia} + \varepsilon_{ia} \tag{2}$$

where $\tilde{T}_{ia}$ is the true cognitive skill of the person and $\varepsilon_{ia}$ is the test measurement error. Similarly, we also only observe $T_{ia^*}$ measured with error.

The presence of measurement error complicates the estimation of age effects even when the errors have mean zero and constant variance. This problem was addressed as early as the inheritance studies of Galton (1889) and has been formalized in Berry et al. (1984). We define the true difference in cognitive skills as:

$$\tilde{\Delta}_a = E(\tilde{T}_{ia^*} - \tilde{T}_{ia}) \tag{3}$$

Assuming that the observed $T_{ia}$ and $T_{ia^*}$ are distributed bivariate normal with a mean of $\mu$ for $T_{ia}$ and a correlation of $\rho$, Berry et al. (1984) derive that:

$$E(T_{ia^*} - T_{ia} | T_{ia}) = \tilde{\Delta}_a - (1-\rho)(T_{ia} - \mu) \tag{4}$$

This relationship shows that for observed cognitive skills below the mean (i.e., $T_{ia} < \mu$), the true difference in skills will tend to be overestimated, while the opposite is true for observations above the mean.

This overall relationship makes intuitive sense. If there is no measurement error (i.e., $\rho = 1$), the difference in observed values equals the true marginal age effect, $\tilde{\Delta}_a$. If, however, there is measurement error, observations far below the mean are likely to be observations that have larger negative measurement errors. These errors are unlikely to be duplicated with similarly large negative errors at the second measurement ($a^*$), meaning that the farther an observation is below the mean, the larger the positive bias in estimated differences. The opposite holds for observations above the mean, for which the estimated differences will be negatively biased. If uncorrected, this phenomenon would bias the observed age-skill patterns whenever true skills differ by age. In practice, estimating the age-skill patterns with unadjusted data yields distorted pictures (see Section 5).

This assessment of the bias – reversion to the mean[5] – yields a natural way to estimate the error-corrected change in cognitive skills from our observed data:

$$\hat{\Delta}_{ia} = (T_{ia^*} - T_{ia}) + (1-r)(T_{ia} - \overline{T}_a) \tag{5}$$

where r and $\overline{T}_a$ are the sample analogs of $\rho$ and $\mu$. This adjustment is used throughout our analysis of patterns in cognitive score changes.[6]

**Age-skill patterns with adjusted data.** Our goals are to employ the individual data to estimate the age-skill pattern and to study how this pattern relates to individual characteristics, behavior, and circumstances. The basic approach is to cumulate the adjusted marginal age changes ($\hat{\Delta}_{ia}$) across the age distribution and to estimate the impact of key individual differences including skill usage and other circumstances across the life cycle.

We can estimate the aggregate age-skill pattern by adding the cumulative averages of (annualized) adjusted marginal changes in skills at each age to the starting skill level at age 16,

---

[5] In various analyses, this phenomenon is also called regression to the mean or simply regression effect.

[6] Equivalently, the same adjustment can be obtained when taking the residual of a regression of the test-score change on the initial test-score level. We also experimented with an alternative adjustment that conditions on a full set of age fixed effects in the regression of skill change on initial skill level, taking out only that part of the variation related to the initial level that is not related to age. The idea is to retain any variation in skill changes that is related to age. Results are nearly identical (not shown).

$$\hat{T}_a = \bar{T}_{i16} + \sum_{\alpha=16}^{a} \bar{\tilde{\Delta}}_\alpha \tag{6}$$

where $\hat{T}_a$ is the adjusted aggregate skill level for individuals with age=$a$ and $\bar{\tilde{\Delta}}_\alpha$ is the average marginal (one-year) change across all individuals with age equal to $a$. Plotting these adjusted scores allows us to produce an adjusted version of Figure 1 based on longitudinal data.[7]

Our development of the full age-skill profile relies on concatenating the marginal age changes across individuals of differing age. As such, it is not immune to possible cohort biases since with the sampling of individuals at a given time respondents of different ages attended school and developed their cognitive skills at different times. The approach implicitly assumes that the development and implications of growing up at different times affects the level of skills but not the rate of change with age over the lifecycle. In other words, if cohort factors such as changes in school quality or motivation toward schooling change the scores for a cohort (i.e., $\tilde{T}_{ia}$ and $\tilde{T}_{ia^*}$) by a constant amount, equation (3) would still yield unbiased estimates of $\tilde{\Delta}_a$.

The subsequent analysis of how age-skill patterns vary with individual backgrounds, behavior, and characteristics can be depicted by relating these factors to the adjusted individual skill data. A stylized version of our basic analytical specification is:

$$\hat{\Delta}_{ia} = \gamma_0 + \gamma_1 a + \gamma_2 a^2 + f(U_{ia}) + g(X_{ia}) + v_{ia} \tag{7}$$

where the $\gamma$ s are parameters of a quadratic age impact on cognitive skills, $f(U_{ia})$ considers usage of skills and $g(X_{ia})$ other systematic influences, and $v$ adds a stochastic term.

These descriptive analyses of observational data do not directly allow for a causal interpretation of the estimated associations. Relating differences in the usage of skills and other characteristics observed at the first testing occasion to subsequent changes in skills shields estimates from outright reverse causation. But the behaviors and characteristics are not distributed randomly, so that they may be associated with unobserved omitted factors that are independently related to skill trajectories. Nonetheless, the descriptive patterns open directions for future research into underlying causal mechanisms.

---

[7] Individuals at the youngest ages are frequently still in school and have skills increasing with age. Considering only changes after 25, when most are not in school, does not change our qualitative results.

## 4. PIAAC Data

Our analysis relies on the unparalleled collection of longitudinal skill data for a representative adult population. The underlying PIAAC assessments include measures of skills that are generally related to life and work outcomes; the samples cover the full adult population; and the data provide information about the background and employment of individuals.

**PIAAC-Longitudinal.** Two waves of repeated testing of the same individuals took place in Germany in 2011/2012 and 2015. The first wave was part of the Programme of the International Assessment of Adult Competencies (PIAAC), an international cross-sectional adult test of literacy and numeracy skills administered by the Organization for Economic Co-operation and Development (OECD) in 39 countries (OECD (2013, 2016)). PIAAC samples are drawn to be representative of the population aged 16-65 in each country. In Germany, the PIAAC-Longitudinal (PIAAC-L) project returned to the initial participants to create a panel study (Rammstedt et al. (2017)).[8] Substantial effort was invested in skill measurement, with an average testing time of roughly 60 minutes and an additional 40 minutes for answering a background questionnaire (Zabal et al. (2014)).[9]

The PIAAC assessment is designed to test skills that are relevant for adults to participate in social life and work situations, as opposed to specific components of cognitive functioning or purely curriculum-based tests. Skills are tested in two main domains, literacy and numeracy.[10] PIAAC defines literacy as the "ability to understand, evaluate, use and engage with written texts in order to participate in society, achieve one's goals, and develop one's knowledge and potential" (OECD (2016)). Numeracy is defined as the "ability to access, use, interpret and communicate mathematical information and ideas in order to engage in and manage the mathematical demands of a range of situations in adult life."

PIAAC-L administered the prior literacy and numeracy tests under conditions identical to the original PIAAC setting. The average time between the first and second test is 3.56 years.

---

[8] Canada, Italy, and Poland also implemented follow-on surveys of the original PIAAC participants, but they did not repeat the skill test (Rammstedt et al. (2017)).

[9] Operationally, data collection was performed under the supervision of trained interviewers in respondents' homes using a computer-based application (with the option of using a paper version) that started with the background questionnaire and followed with the skill test.

[10] The original PIAAC assessment also included the optional (to countries and individuals) domain of 'problem solving in technology-rich environments' for the subsample of participants with confident computer usage, but this domain was not included in PIAAC-L.

From the 5,379 initial participants from 2012, a retaker sample of 3,263 participants (60.7 percent) was tested again in 2015 (Rammstedt et al. (2017), Zabal, Martin, and Rammstedt (2017)). An analysis of the original and retaker samples does not indicate that selection issues are likely to bias our age-skill estimates. Comparing the cross-sectional age-skill profiles for the two samples using the original sampling weights (Panel A of Appendix Figure A2), retakers are slightly positively selected in terms of achievement, a phenomenon generally consistent with other longitudinal assessment surveys (Martin et al. (2021)). Reassuringly for our purposes, however, this positive selection does not differ by age, with observed age-skill profiles shifted upwards in parallel. Importantly, the differences disappear when using the PIAAC-L sampling weights to adjust the retaker sample to the general population in terms of observables (Panel B). Thus, the sampling weights (used throughout our analysis) ensure depictions that are representative for the full population.

To allow for joint analysis of achievement on the two test occasions, PIAAC-L uses Item Response Theory (IRT) scaling procedures to provide ten plausible values of achievement measures for both waves. We use all ten plausible values throughout our analysis, clustering standard errors at the level of individuals. For analytical purposes, we standardize scores to have mean zero and standard deviation (SD) one in the 2012 test for the retaker sample.

The average age in our analysis sample is 41.3 years (Appendix Table A1). 37.8 percent have a white-collar occupation,[11] and 30.6 percent have a tertiary education. A first look at the intertemporal test data indicates that literacy scores are on average 0.049 SD higher and numeracy scores are 0.020 SD higher on the second test compared to the first test. These positive average changes, observed within individuals, cast doubt that the overall negative slopes of the cross-sectional age-skill profiles in Figure 1 mostly reflect actual age effects.[12] Most of our analysis focuses on the sample of employed (2,497 observations) for whom we can observe skill usage patterns at both work and home. For this sample, average achievement as well as shares of tertiary education and white-collar occupations are somewhat higher. Nonetheless, the overall age-skill patterns we analyze follow very similar patterns as those in the full sample.[13]

---

[11] White-collar occupations are measured by codes 1-4 of the International Standard Classification of Occupations (ISCO): managers, professionals, technicians and associate professionals, and clerical support workers.

[12] In our setting, it is very unlikely that observed skill changes are affected by learning from past test responses (Salthouse (2010a)) because: (a) the long time distance of 3-4 years between the two tests makes it very unlikely that participants remember much from the first test when taking the second round; (b) most participants will not get identical test items because of the rotation of test booklets that underlies the IRT scaling; and (c) participants are neither informed about their scores nor receive feedback on the correctness of their answers in the first test.

**Skill usage.** In addition to measuring skills, the initial PIAAC wave collects rich information on individuals' usage of skills. The background questionnaire includes separate item batteries on the frequency of respondents' activities related to reading and to math at work and in everyday life. The reading usage items refer to the frequency of reading various types of documents,[14] and the math usage items refer to the frequency of various activities involving numbers and math.[15] In both domains, the usage measures that we construct contain six items, and the same items are used for skill usage on the job and in everyday life. The usage frequency for each is measured on a five-point scale from never to every day. We collapse the separate categories into a composite measure by detecting whether individuals perform each given activity at least once a month and then taking a simple average over the six work items and the six home items of reading and math usage, respectively.[16]

The usage of skills varies widely by domain and background characteristics. On average, respondents perform 59.7 percent of the reading activities and 35.3 percent of the math activities at least once per month. Workers in white-collar occupations use skills much more frequently than workers in blue-collar occupations (69.7 percent vs. 49.7 percent in reading and 43.2 percent vs. 27.3 percent in math); see Appendix Figure A4 for details. Similar differences exist between respondents with and without tertiary education. While there is no gender difference in reading usage, men perform math activities more frequently than women (38.2 percent vs. 32.0 percent). Interestingly, most of these usage patterns do not differ strongly between workers above or below the sample median of age (Appendix Figure A4). The main exception is that the math usage of blue-collar workers and of individuals without tertiary education declines with age. Interestingly, overall reading usage is more frequent at home than at work (65.2 percent vs. 54.2 percent), whereas the opposite is true for math usage (32.8 percent vs. 37.7 percent). The correlation between skill usage at work and at home is 0.411 for reading and 0.343 for math. Reading and math usage are correlated at 0.506. Skill usage is also significantly correlated with test scores (0.357 for reading usage and literacy and 0.401 for math usage and numeracy).

---

[13] See Section 5 below. The cross-sectional age-skill pattern in the employed sample indicates a decline in literacy starting in the early thirties and in numeracy starting in the early forties (Appendix Figure A3).

[14] The six reading items refer to how often respondents usually 1. read directions or instructions; 2. read letters, memos or e-mails; 3. read articles in newspapers, magazines or newsletters; 4. read articles in professional journals or scholarly publications; 5. read books; and 6. read manuals or reference materials. There are two additional items in this battery – read bills, invoices, bank statements or other financial statements; and read diagrams, maps or schematics – that may reflect math rather than reading usage. We do not include these items in our literacy usage measure, but qualitative results are identical if we do (not shown). In addition, there is a separate battery of four items on writing activities. If we combine these items with the reading items, qualitative results are again the same.

[15] The six math items refer to how often respondents usually 1. calculate prices, costs or budgets; 2. use or calculate fractions, decimals or percentages; 3. use a calculator; 4. prepare charts, graphs or tables; 5. use simple algebra or formulas; and 6. use more advanced math or statistics (calculus, trigonometry, regressions).

[16] Results are similar when linearizing across the five frequency-of-use categories rather than focusing on at least monthly usage (see Section 5). The publicly available PIAAC database contains derived measures of skill usage, but its specific approach leads to a highly biased sample for the usage measures. The PIAAC measures code all respondents who respond "never" to the usage frequency for all activities in a battery as missing (despite having an active response as opposed to refused answers or "do not know" answers). The reasons given for this procedure are that the chosen derivation approach would suffer from zero-inflated-count issues and would not be comparable to another measure of usage of ICT skills which is elicited only among respondents who report to use computers.

## 5. Results on Overall Age-Skill Profiles

The changes in skills observed within individuals over time show a very different pattern than suggested by the cross-sectional data. Panel B of Figure 2 depicts the annualized marginal changes in scores with age ($\overline{\hat{\Delta}_\alpha}$) for individuals at each age for the full population sample (3,263 observations). The analysis corrects for reversion to the mean according to equation (5) and includes a quadratic fit and 95-percent confidence intervals. The quadratic prediction indicates increases in skills for individuals up to age 45 in literacy and up to age 40 in numeracy. Skill changes turn negative beyond these ages, with declines notably stronger in numeracy than in literacy. In both subjects, the marginal change in skills declines steadily with age, and the decline is initially close to linear and ultimately slows slightly.[17]

The longitudinal age-skill profiles implied by these individual changes indicate substantial increases in skills from early adult life into the early forties, followed by modest declines in literacy and more significant declines in numeracy. Panel A of Figure 2 displays the cumulative age-skill profiles calculated according to equation (6). Literacy skills increase strongly in the twenties and thirties and tend to stabilize and flatten starting in the late thirties. The peak is at age 46, but subsequent declines are limited. Numeracy skills also increase strongly at young ages but peak earlier, at age 41, and decline significantly at later ages, although not below the levels observed in the early twenties.

---

[17] Skill changes with age and age-skill profiles look quite similar when estimated just on the sample of individuals aged 25 and older (Appendix Figure A5), implying that the full-sample results are not driven by patterns in the age group 16-25 where many individuals are still in education.

These age patterns contrast sharply with the cross-sectional patterns of Figure 1 which indicated ubiquitous declines. The difference in the age patterns across subjects is particularly surprising given that achievement in the two subjects is strongly correlated in the cross-section (correlation of 0.86), but it is consistent with the subject-specific aggregate patterns of skill usage described in Section 3. We expand on this in the next section.

## 6. Results on Age-skill Patterns by Skill Usage and Background

Previous studies frequently suggest considerable heterogeneity in age-skill patterns across individuals with different backgrounds. We extend these studies by explicitly considering how skill trajectories relate to individuals' usage of skills. For these analyses, we focus on the sample of employed workers (2,497 observations) for whom we can observe occupations and skill usage patterns at work and at home.[18]

**Heterogeneity by usage.** When we separate age-skill profiles by the frequency of skill usage, we get strikingly different pictures. We repeat our prior analysis for samples divided at the median of the aggregate measure of skill usage at work and at home derived from the initial survey.[19] Strikingly, those with above-median usage of the respective skill on average never show a decline in skills (Figure 3).[20] Their skills increase steeply into the fifties and then flatten out, with no indication of average decline. By contrast, for those with below-median usage, skill decline begins in their mid-thirties. Interestingly, in contrast to the aggregate pattern, the usage-specific patterns are quite similar for the two skill dimensions.[21] These results are consistent with skill usage playing a leading role in determining whether skills are gained, retained, or lost over time.

The findings about age-skill patterns for the low- and high-usage groups in Figure 3 also vividly demonstrate the importance of adjustment for reversion to the mean. The key differences in age-skill patterns with usage are distorted when looking at the raw, unadjusted skill data. The high-usage group has above-the-mean average test performance, implying changes will be biased downward by the error pattern if uncorrected. With uncorrected data, the high-usage group appears to have lower growth in literacy skills with age and to lose numeracy skills after age 40 (Appendix Figure A11). On the other hand, for both skill categories the low-usage group appears to show much lower declines in skills at older age – the kind of positive bias that reversion to the mean causes for observations with initially low scores. Looking at these patterns with unadjusted data would thus change the conclusions about age-skill patterns.

---

[18] The overall age-skill profiles for the employed sample (Appendix Figure A6) look very similar to the one for the full population (Figure 2).

[19] To account for initial differences in levels between subgroups in the figures, we predict scores at age 16 from a regression of scores on a quadratic in age within the respective subgroup in the initial wave.

[20] Again, the pattern is qualitatively unchanged when disregarding individuals younger than 25 in the analysis (Appendix Figure A7).

[21] Results look qualitatively similar when considering skill usage at work and at home separately (Appendix Figures A8 and A9). Skill usage at home is observed not just in the sample of employed, but in the full population sample; full-population results look very similar, as well (Appendix Figure A10).

**Heterogeneity by background characteristics.** Prior analyses, lacking direct usage data, have focused on how patterns differ across readily observed subgroups – occupations, education levels, and gender. These analyses are typically motivated by assumed average usage differences, assumptions that are indeed validated in our data but that ignore within-group usage differences.

On average, the substantially higher skill usage of white-collar and tertiary-educated workers (see Appendix Figure A4) shows up in skill patterns. Distinguishing between blue-collar and white-collar occupations yields results that are broadly similar to the split between low- and high-usage individuals. Skills consistently increase throughout the age range for white-collar workers but start to decline early for blue-collar workers (Panel A of Figure 4).[22] The increase for white-collar workers is more pronounced in literacy than in numeracy, particularly at older ages. Virtually the same pattern is observed when distinguishing between those with and without a tertiary education (Panel B).

For gender, patterns differ somewhat between the two subjects. In literacy, skill trajectories are initially similar across genders but then flatten out for men and decline slightly from the mid-forties for women (Panel C of Figure 4). The numeracy pattern is similar for men, whereas for women, numeracy skills start to decline from their early thirties and do so more strongly. The more pronounced gender differences in numeracy are consistent with the more frequent math usage of men, whereas there are no gender differences in reading usage (see Appendix Figure A4). We return to this below.[23]

---

[22] Appendix Figures A12-A14 show the underlying analyses of marginal skill changes by age for the three subgroup divisions.

[23] We can also distinguish by the initial level of test scores. Increases at younger ages are quite similar for initially low- and high-achieving individuals, followed by a flattening at older ages for initially high-achieving individuals and some decrease for initially low-achieving individuals (Appendix Figure A15).

To summarize the aggregate difference in profiles by subgroup, we perform regression analyses that allow for multiple influences simultaneously (Table 1). Following equation (7), the regressions express the change in skills ($\hat{\Delta}_{ia}$) as a quadratic function of age and approximate the subgroup effects by linear shifts in the skill changes for all ages. Going from no monthly usage of skills to monthly usage in all categories is associated with an average increase in the annual change in skills by a statistically highly significant 0.108 SD in literacy and 0.100 SD in numeracy on average (columns 1 and 6).[24] The average difference in skill changes between blue- and white-collar workers is 0.056 SD in literacy and 0.030 SD in numeracy (columns 2 and 7). The age-skill pattern is shifted by 0.057 SD in literacy and 0.039 SD in numeracy between workers with and without tertiary education (columns 3 and 8) and by 0.014 SD in literacy and 0.026 SD in numeracy between men and women (columns 4 and 9). Coefficient estimates are mostly reduced in size when the full set of measures is jointly considered, but they generally maintain their relative ordering and their significance (columns 5 and 10). The main exception is that the coefficient on white-collar occupations in numeracy becomes small and loses significance.[25]

**Post-40 skill trajectories by characteristics and usage.** As suggested, interpreting the age-skill patterns across the background factors is complicated by their interactions with varying amounts of skill usage (see Appendix Figure A4). To highlight these interactions, we plot the average change in skills in various subgroups for the population over age 40, roughly above the population mean.[26] If there are different aging patterns, the previous analysis suggests that they would typically be apparent at these ages.

Within each of the background subgroups, there are large differences in skill trajectories by whether individuals do or do not frequently use the respective skills Figure 5. Even among those groups with average skill growth by age – white-collar and tertiary-educated workers – only individuals with frequent skill usage show increases in skills beyond age 40. By contrast, individuals in these groups with below-average usage show no significant changes. Similarly, in the groups previously showing average age-related declines – blue-collar and less educated workers – those with above-average usage do not show significant skill losses beyond age 40 (except for blue-collar workers in literacy). In sum, even within these population subgroups marked by generally different positions in the economy, whether individuals gain or lose skills in the second half of their working-age life is strongly mediated by their frequency of skill usage.

---

[24] Results are robust to using various alternative measures of skill usage (Appendix Table A2). Skill usage at work and skill usage at home are significant when entered individually (columns 1-2 and 7-8). When included jointly, both remain significant, but the point estimates for usage at work are twice as large as for usage at home (columns 3 and 9). When reading and math usage are included jointly, math usage is insignificant in the literacy regression, but reading usage is significant in the numeracy regression (columns 4 and 10).

[25] The skill-usage measures consider only whether an activity is performed at least once a month. Alternatively, we can use the full information of the underlying five-point scale that ranges from never to every day, linearizing the five options from zero to one. Coefficients on this alternative usage measure are even larger in both subjects (Appendix Table A2, columns 5-6 and 11-12).

[26] Qualitative results are similar in the smaller subsamples of workers aged over 45 or 50 (not shown).

The distinction by gender follows a somewhat different pattern. In literacy, usage dominates the overall age-skill relationship for both males and females. But in numeracy, usage differences have noticeably smaller differential effects on age patterns for females compared to males. This pattern is not driven by differences in the subgroup shares falling above or below the overall sample median of usage, as they emerge similarly when splitting the subgroups by the median in their respective subgroup (Appendix Figure A16). Instead, the age trajectory of numeracy skills for women appears to be less dependent on usage.

## 7. Conclusions

Declining cognitive skill with age is quite commonly treated as inevitable. Much of the existing research supports such a conclusion. We find that this conclusion is subject to serious data and analytical issues. Little of the prior research has been able to track the actual cognitive skill changes of individuals in representative populations. Instead, it either relied on observing group differences in skills by individuals of different ages or on skill changes in unrepresentative samples. As such, the first approach is naturally not able to disentangle individual aging patterns from underlying cohort differences, while it is difficult to know how to generalize from the second approach with selected samples.

Data for a representative sample of the German adult population permit tracing the skill changes that are seen with aging. Moreover, because the data from the PIAAC survey of the OECD contain not only repeated cognitive skill assessments but also survey information about individuals' reading and math activities at work and at home, it is possible to link individual skill changes to differences in skill usage.

Cross-sectional patterns show decreasing average skills starting early in adulthood. By contrast, moving to age-skill patterns based on longitudinal data and skill measures corrected for reversion to the mean leads to the conclusion that average skills increase through middle age. Literacy skills decline little from their peak in the mid-forties. Numeracy skills show a sharper decrease after the early forties but do not decline below levels in the early twenties.

Heterogeneity in the population based on individual behavior is the most striking aspect of the age-skill patterns. Especially important is skill usage at different ages over the life cycle. Individuals who have above-average skill usage at work or at home forestall any decline and continue to gain literacy and numeracy skills at older ages. Consistent with the aggregate patterns of skill usage, these patterns show up in the skill evolution by occupation and education that themselves are consistent with differential skill usage. Below-average usage by either white-collar workers or college graduates is still associated with stagnating skills with age.

While the skill trajectories of men flatten in their forties, the literacy trajectory of women – which is broadly parallel to men into the early forties – declines somewhat at older age, and the numeracy trajectory of women declines earlier and more strongly. Noticeably, the impact of usage on the age pattern of numeracy skills is more limited for women.

In general, the strong differences in age-skill profiles found in our analysis for different skill usage and background characteristics imply that considerable caution is needed in interpreting analyses of unrepresentative samples. The selected samples entering prior conclusions about age-skill patterns are unlikely to generalize to the overall population.

Our analysis of these questions was motivated by the rapid aging of many advanced economies. The cognitive skills of the population are important not only for individual incomes but also for the economic growth of nations. Overall, our results are not consistent with a view that a natural law dictates an inevitable decline in skills with age. This is consolation for countries with aging populations, but avoidance of skill losses is not automatic and appears related to stimulation from skill usage. These results thus suggest that age-skill relationships of adults might deserve policy attention.

**Figure 1: Cross-sectional age-skill profiles**

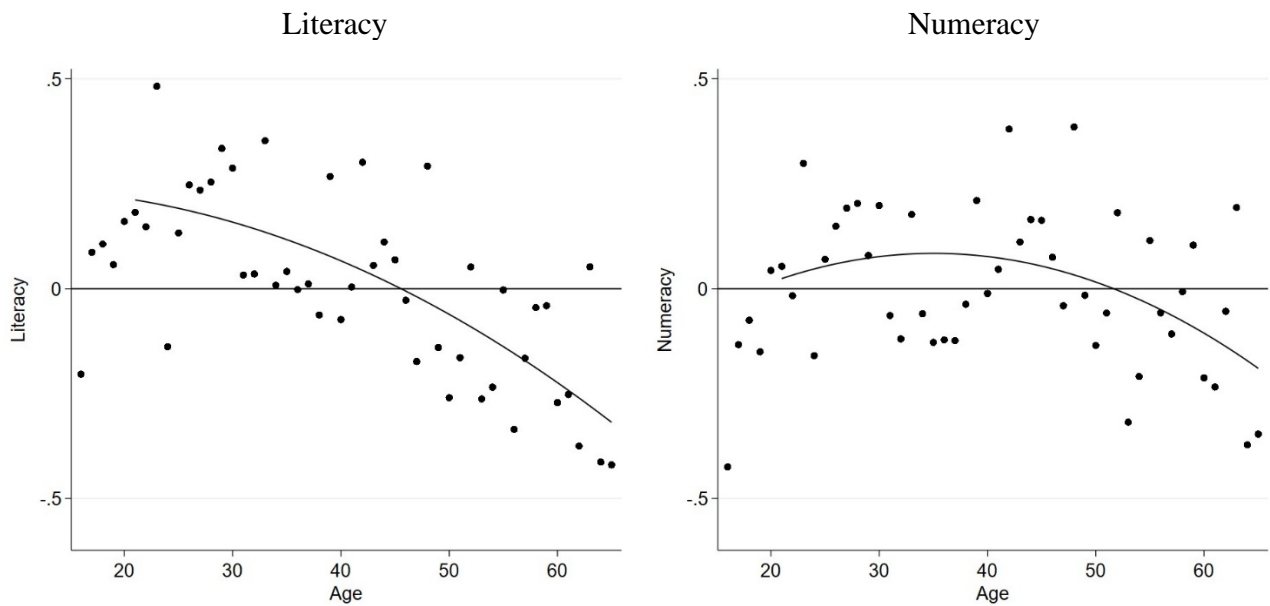

Notes: Cross-sectional association between age and skills in the initial (2012) wave. Dots: average skills by age. Line: quadratic fit (estimated over 21-65 age range). Sample: full population, ages 16-65, weighted by sampling weights (N = 3,263). Data source: PIAAC-L.

**Figure 2: Longitudinal age-skill profiles**

A. Cumulative age-skill profiles

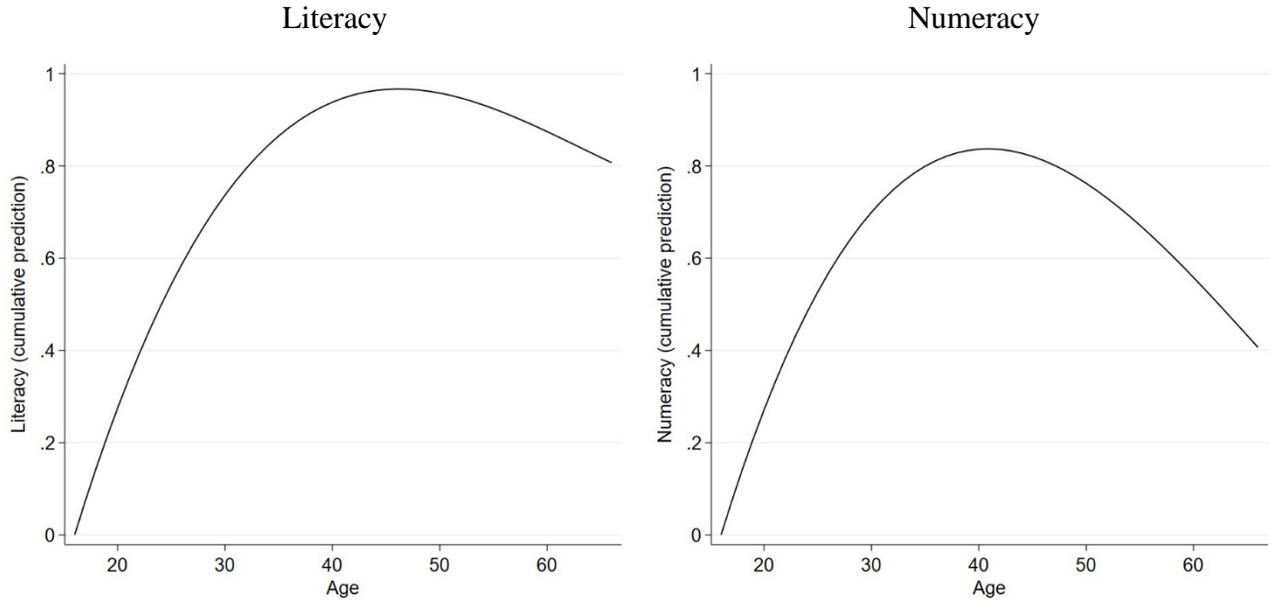

B. Marginal skill changes by age

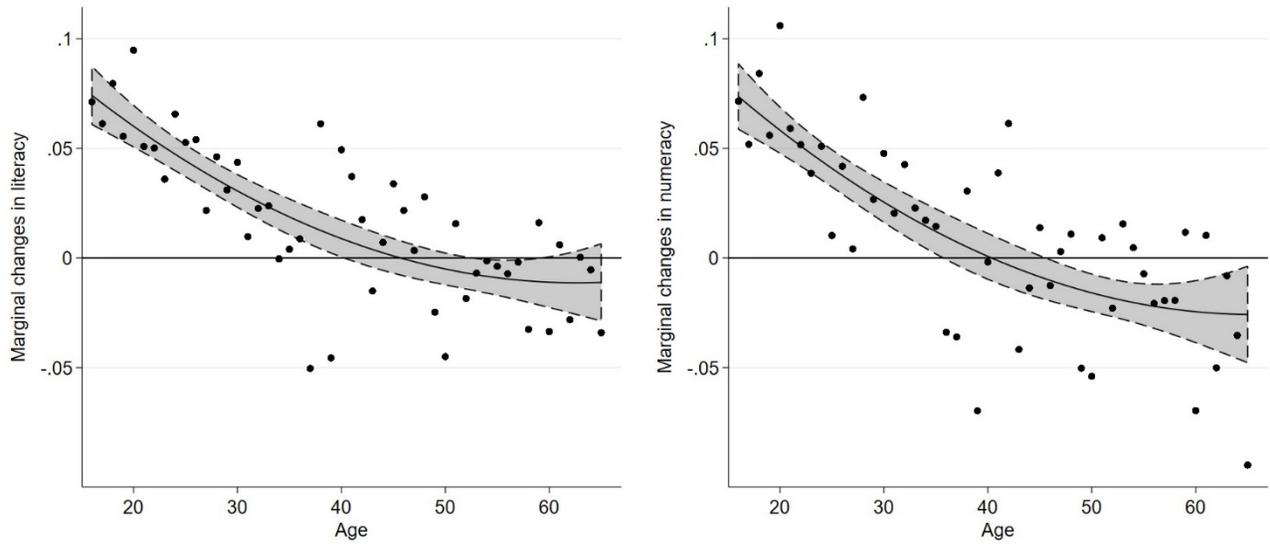

Notes: Panel A: cumulative depiction of the predicted marginal change in skills at each age. Panel B: marginal annualized change in skills between the two waves by age, adjusted for reversion to the mean. Dots: average individual marginal change in skills by age. Line: quadratic fit. Gray area: 95 percent confidence interval. Sample: full population, ages 16-65, weighted by sampling weights (N = 3,263). Data source: PIAAC-L.

**Figure 3: Age-skill profiles by skill usage**

A. Cumulative age-skill profiles

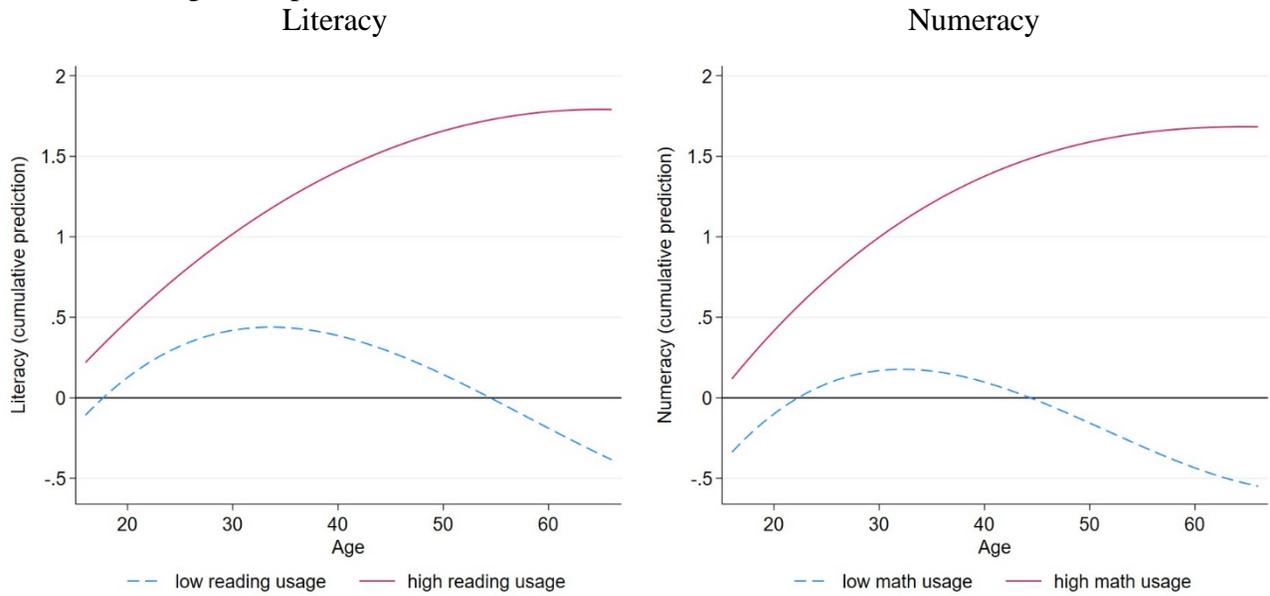

B. Marginal skill changes by age

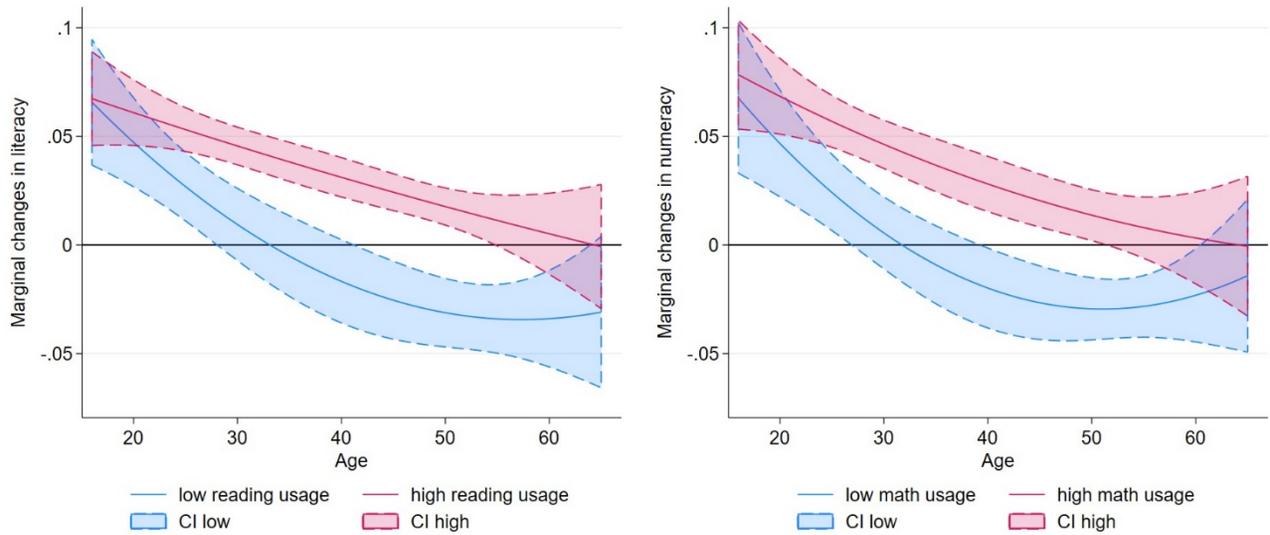

Notes: Panel A: cumulative depiction of the predicted marginal change in skills at each age. Panel B: quadratic fit (with 95 percent confidence interval (CI)) of marginal annualized change in skills between the two waves by age, adjusted for reversion to the mean. Sample split by median of skill usage at work and at home. Sample: employed workers, ages 16-65, weighted by sampling weights (N = 2,497). Data source: PIAAC-L.

**Figure 4: Age-skill profiles by background characteristics**

A. By occupation

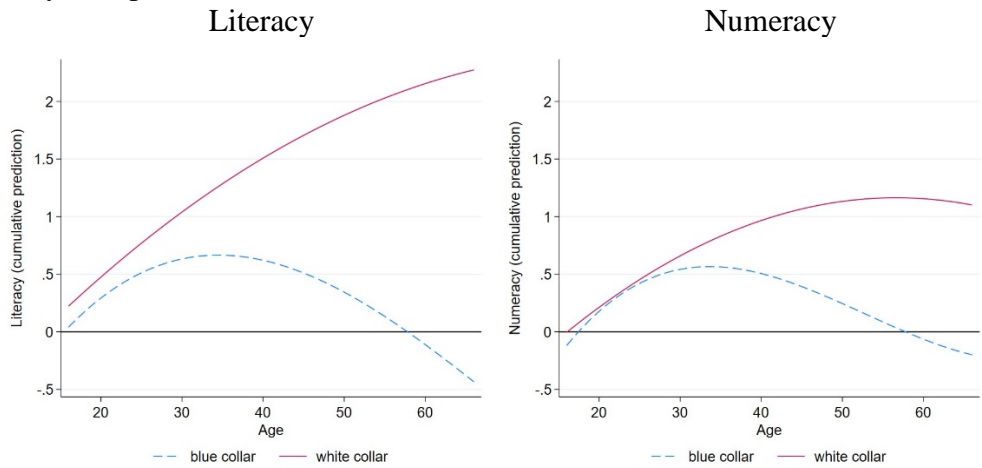

B. By education

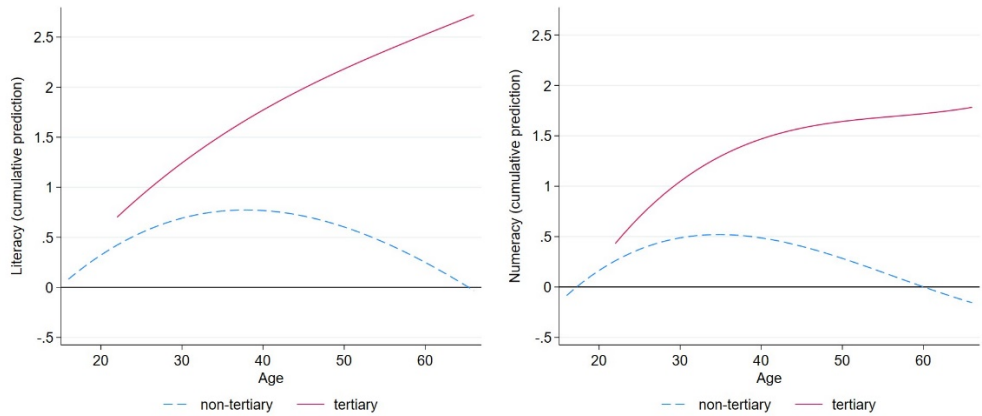

C. By gender

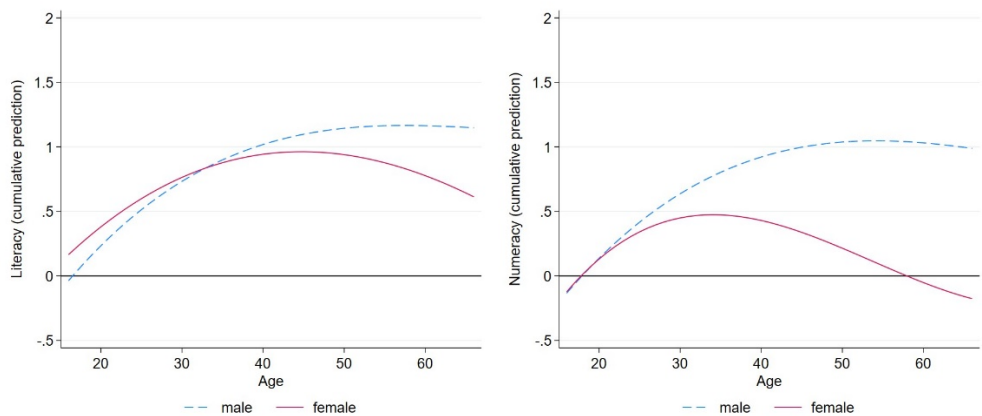

Notes: Cumulative depiction of the predicted marginal annualized change in skills between the two waves at each age, adjusted for reversion to the mean. Sample splits by blue-/white-collar occupations, (no) tertiary education, and gender, respectively. Sample: employed workers, ages 16-65, weighted by sampling weights (N = 2,497) (ages 22-65 for completed tertiary education). Data source: PIAAC-L.

**Figure 5: Skill changes after age 40: By background characteristics and skill usage**

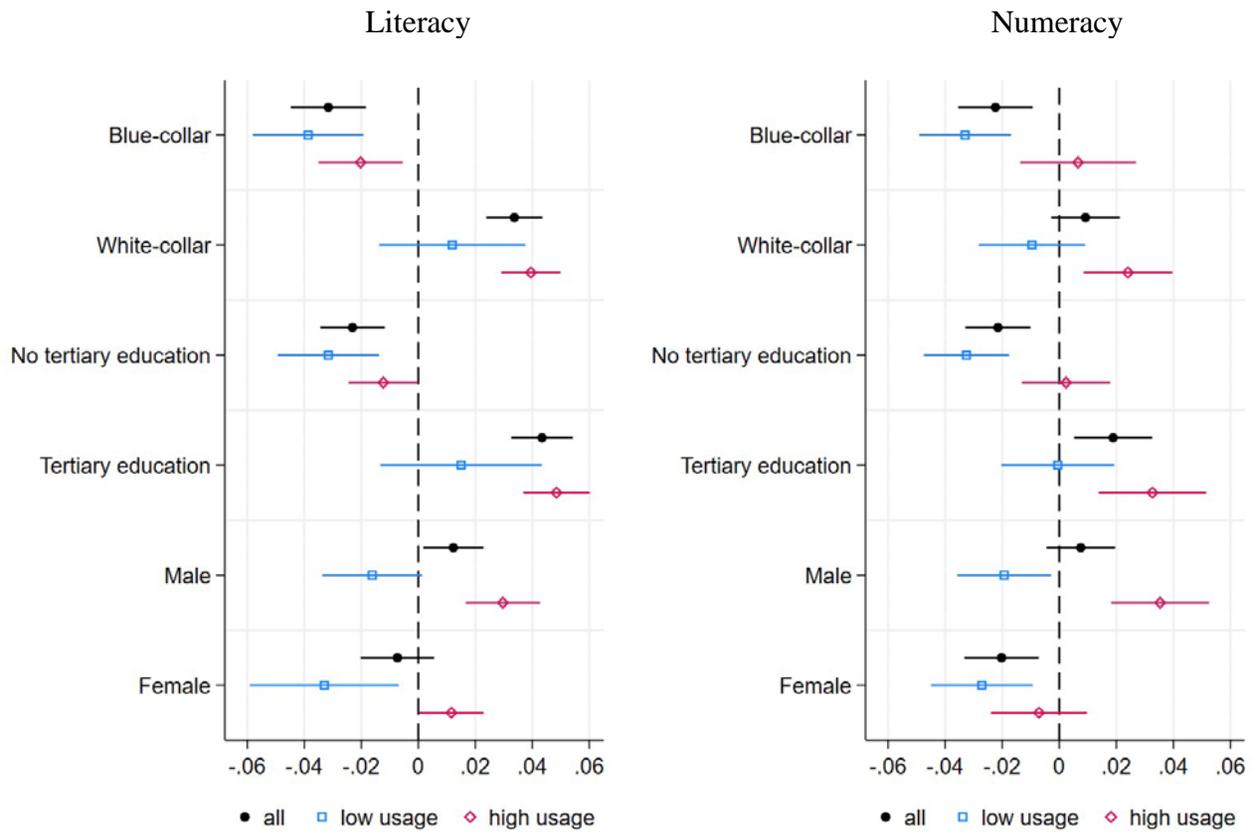

Notes: Average individual marginal annualized change in skills between the two waves, adjusted for reversion to the mean, and 95 percent confidence band. Subgroup means by blue-/white-collar occupations, (no) tertiary education, and gender, respectively. Low/high skill usage: below/above median of skill usage at work and at home. Sample: employed workers, ages 40-65, weighted by sampling weights. Data source: PIAAC-L.

**Table 1: Heterogeneity in marginal changes in skills**

|  | Literacy | | | | | Numeracy | | | | |
|---|---|---|---|---|---|---|---|---|---|---|
|  | (1) | (2) | (3) | (4) | (5) | (6) | (7) | (8) | (9) | (10) |
| Age | -0.0042*** | -0.0047*** | -0.0052*** | -0.0033** | -0.0055*** | -0.0059*** | -0.0063*** | -0.0069*** | -0.0054*** | -0.0067*** |
|  | (0.0015) | (0.0015) | (0.0015) | (0.0015) | (0.0015) | (0.0018) | (0.0019) | (0.0018) | (0.0018) | (0.0018) |
| Age squared | 0.030 | 0.035* | 0.039** | 0.020 | 0.043** | 0.054** | 0.055** | 0.060*** | 0.046** | 0.062*** |
|  | (0.019) | (0.019) | (0.019) | (0.019) | (0.019) | (0.023) | (0.024) | (0.023) | (0.023) | (0.023) |
| Skill usage | 0.108*** |  |  |  | 0.056*** | 0.100*** |  |  |  | 0.074*** |
|  | (0.015) |  |  |  | (0.016) | (0.018) |  |  |  | (0.018) |
| White-collar occupation |  | 0.056*** |  |  | 0.034*** |  | 0.030*** |  |  | 0.011 |
|  |  | (0.006) |  |  | (0.007) |  | (0.008) |  |  | (0.010) |
| Tertiary education |  |  | 0.057*** |  | 0.030*** |  |  | 0.039*** |  | 0.024** |
|  |  |  | (0.007) |  | (0.008) |  |  | (0.009) |  | (0.012) |
| Female |  |  |  | -0.014** | -0.017*** |  |  |  | -0.026*** | -0.022*** |
|  |  |  |  | (0.006) | (0.006) |  |  |  | (0.008) | (0.008) |
| Constant | 0.065** | 0.116*** | 0.137*** | 0.119*** | 0.108*** | 0.115*** | 0.151*** | 0.165*** | 0.159*** | 0.142*** |
|  | (0.027) | (0.027) | (0.027) | (0.028) | (0.027) | (0.033) | (0.033) | (0.033) | (0.034) | (0.033) |
| $R^2$ (adj.) | 0.046 | 0.050 | 0.048 | 0.020 | 0.070 | 0.036 | 0.025 | 0.029 | 0.023 | 0.046 |
| Observations | 2,497 | 2,497 | 2,497 | 2,497 | 2,497 | 2,497 | 2,497 | 2,497 | 2,497 | 2,497 |

Notes: Least squares regressions weighted by sampling weights. Dependent variable: individual marginal annualized change in skills between the two waves, adjusted for reversion to the mean. Age squared divided by 1,000. Skill usage: average of indicators of at least monthly skill usage of battery of usage categories at work and at home. Sample: employed workers, ages 16-65. Regressions use ten plausible values of skill measurement per observation (individual). Standard errors clustered at the individual level in parentheses. Significance level: *** 1 percent, ** 5 percent, * 10 percent. Data source: PIAAC-L.

**Figure A1: Cross-sectional age-skill profiles: OECD countries**

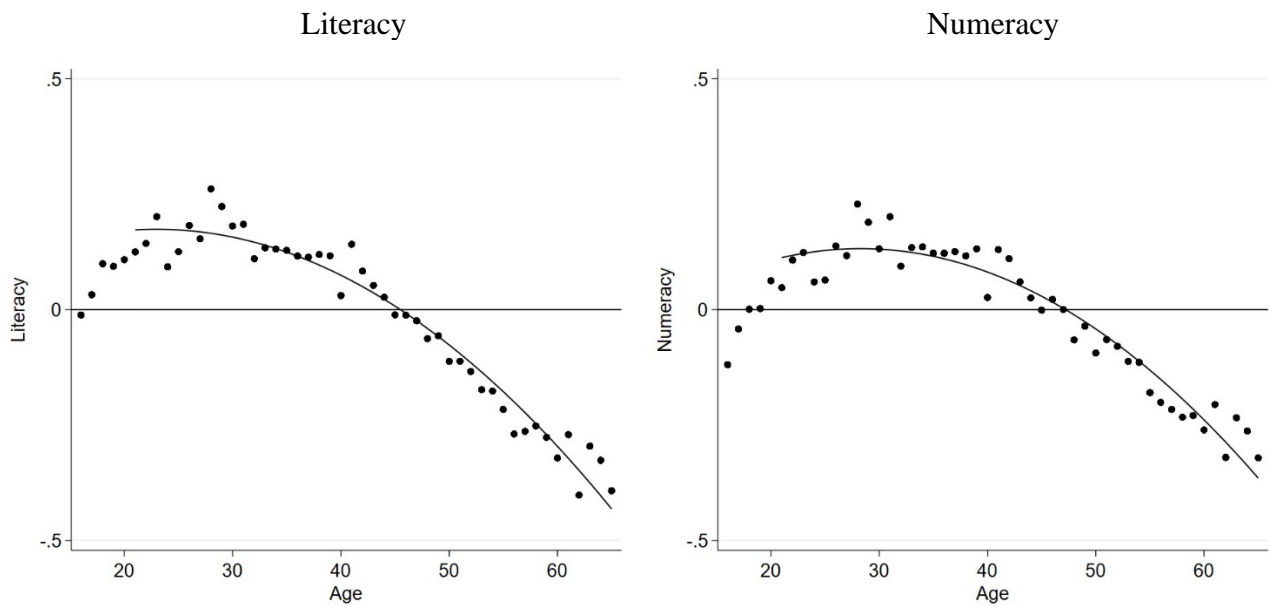

Notes: Cross-sectional association between age and skills in PIAAC (2012). Dots: average skills by age. Line: quadratic fit (estimated over 21-65 age range). Sample: 25 OECD countries with continuous age data; full population, ages 16-65, weighted by sampling weights (N = 147,667). Data source: PIAAC.

**Figure A2: Cross-sectional age-skill profiles: Original vs. retaker sample**

A. Using original weights

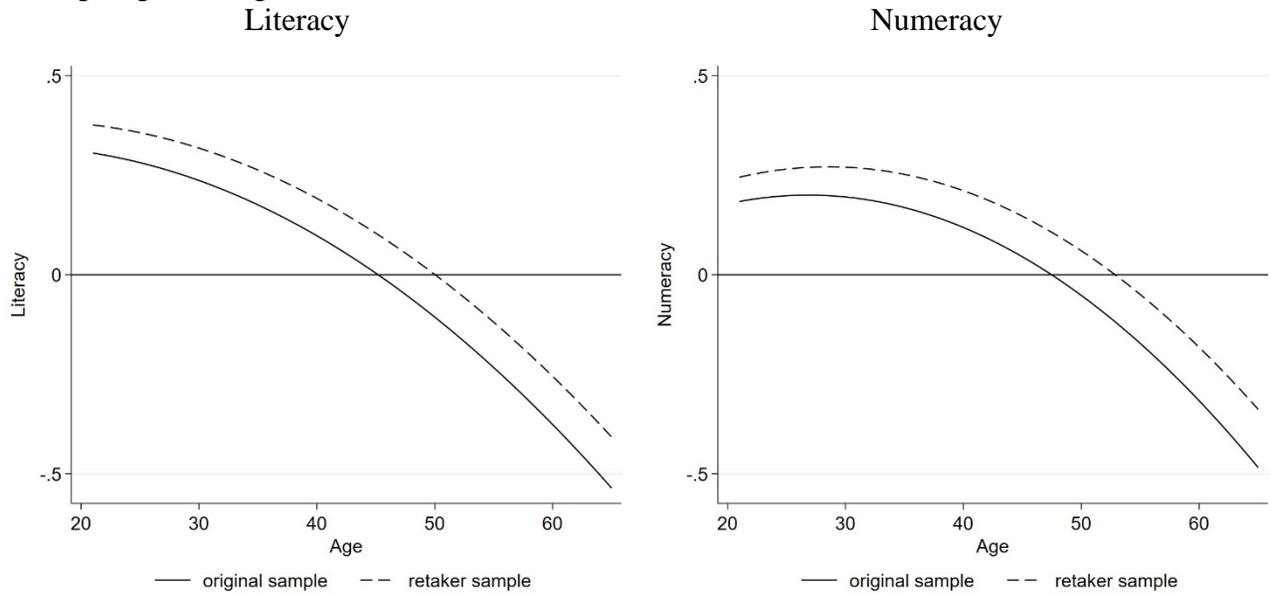

B. Using new sampling weights for retaker sample

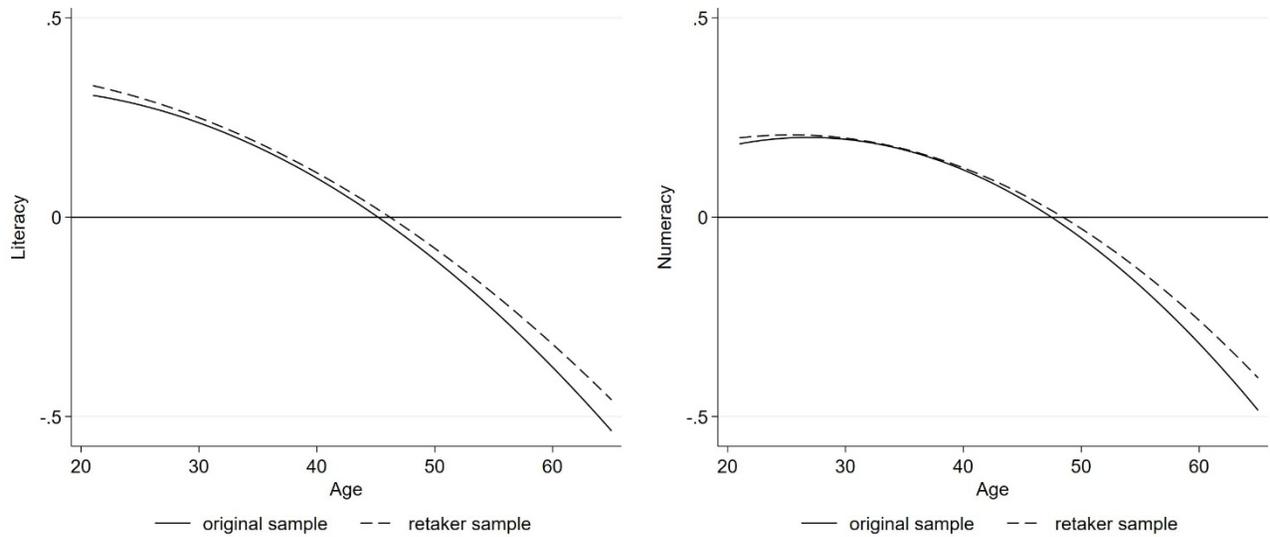

Notes: Quadratic fit of the cross-sectional association between age and skills in the initial (2012) wave, estimated over 21-65 age range. Original IRT scaling. Sample: full population, ages 16-65, weighted by sampling weights. Original sample: N= 5,379. Retaker sample: N = 3,263. Data source: PIAAC-L.

**Figure A3: Cross-sectional age-skill profiles: Employed sample**

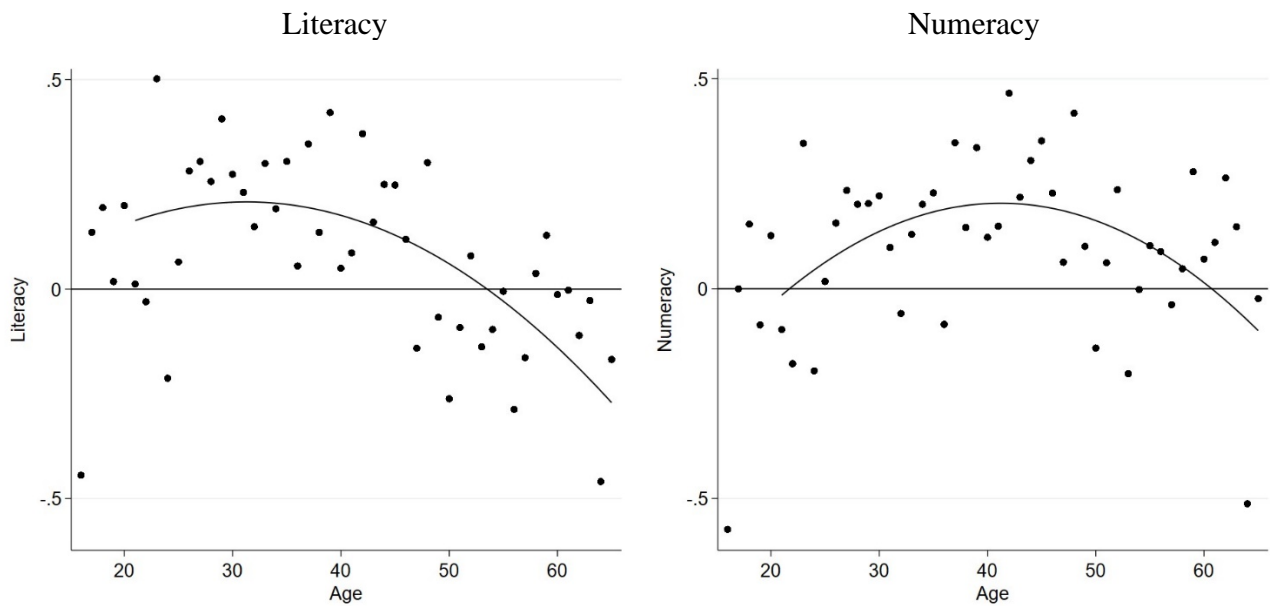

Notes: Cross-sectional association between age and skills in the initial (2012) wave. Dots: average skills by age. Line: quadratic fit (estimated over 21-65 age range). Sample: employed workers, ages 16-65, weighted by sampling weights (N = 2,497). Data source: PIAAC-L.

**Figure A4: Skill usage: By background characteristics and age**

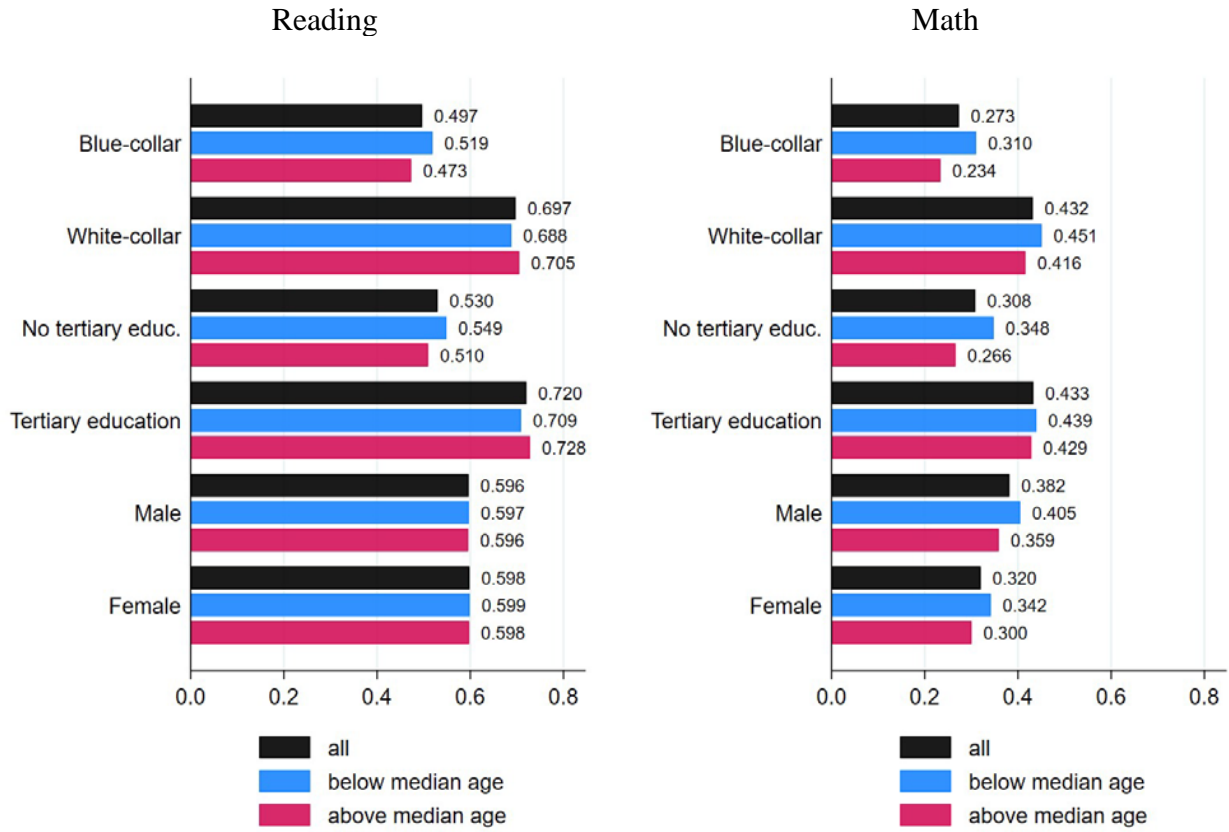

Notes: Average of indicators of at least monthly skill usage of battery of usage categories at work and at home. Subgroup means by blue-/white-collar occupations, (no) tertiary education, and gender, respectively. Below/above median age: sample split by median of age (43). Sample: employed workers, ages 16-65, weighted by sampling weights. Data source: PIAAC-L.

**Figure A5: Longitudinal age-skill profiles: Sample aged 25-65**

A. Cumulative age-skill profiles

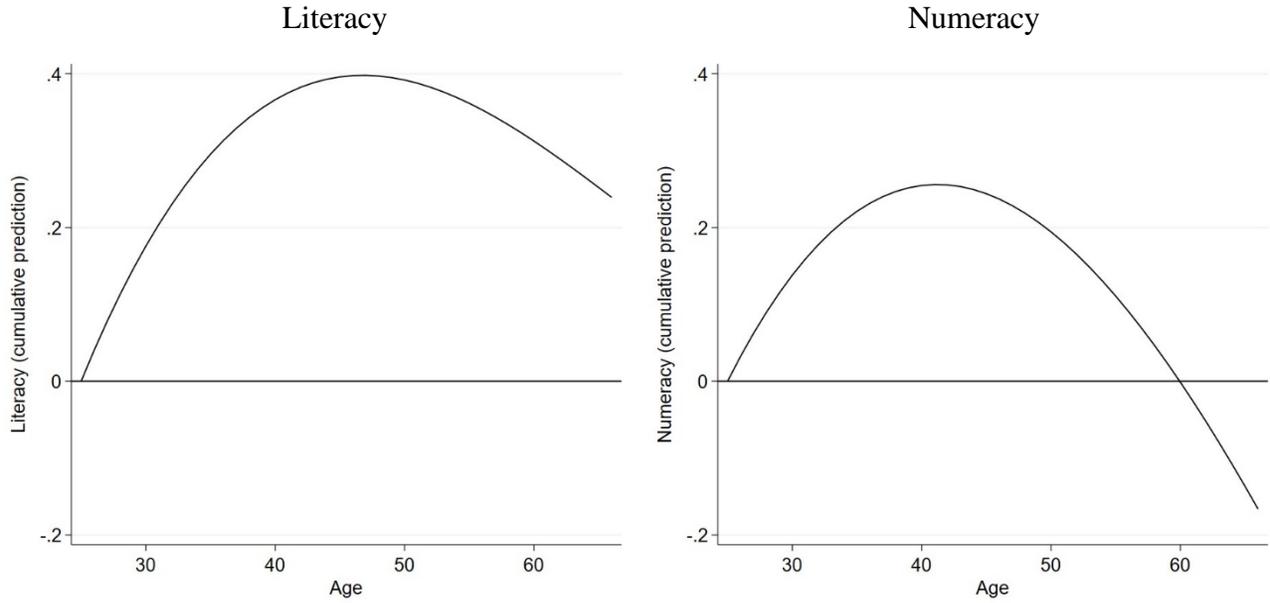

B. Marginal skill changes by age

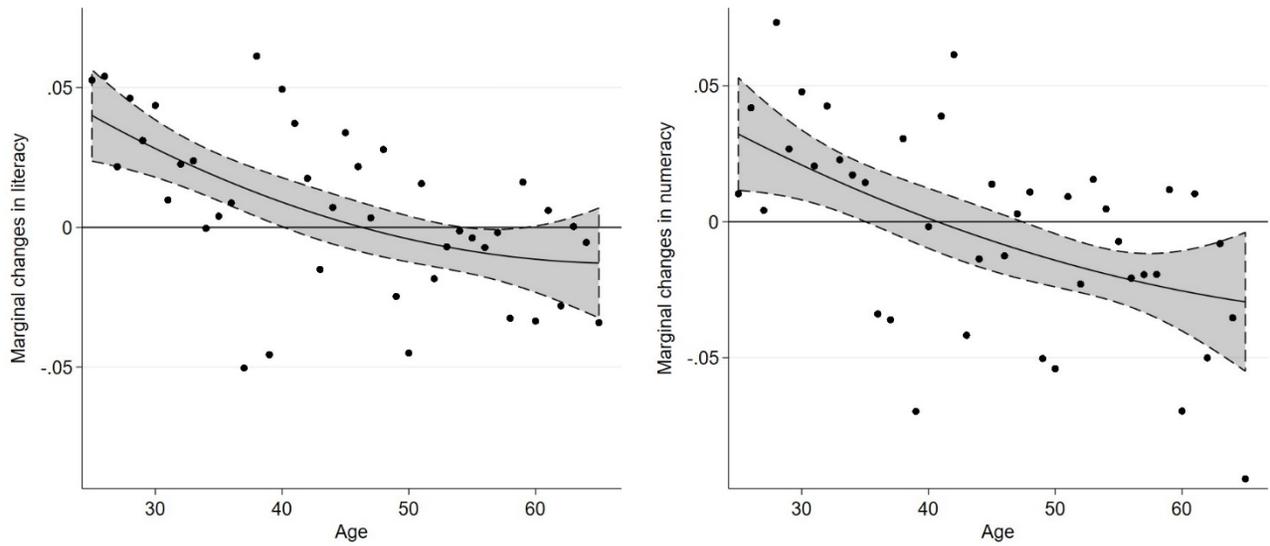

Notes: Panel A: cumulative depiction of the predicted marginal change in skills at each age. Panel B: marginal annualized change in skills between the two waves by age, adjusted for reversion to the mean. Dots: average individual marginal change in skills by age. Line: quadratic fit. Gray area: 95 percent confidence interval. Sample: full population, ages 25-65, weighted by sampling weights (N = 2,664). Data source: PIAAC-L.

**Figure A6: Longitudinal age-skill profiles: Employed sample**

A. Cumulative age-skill profiles

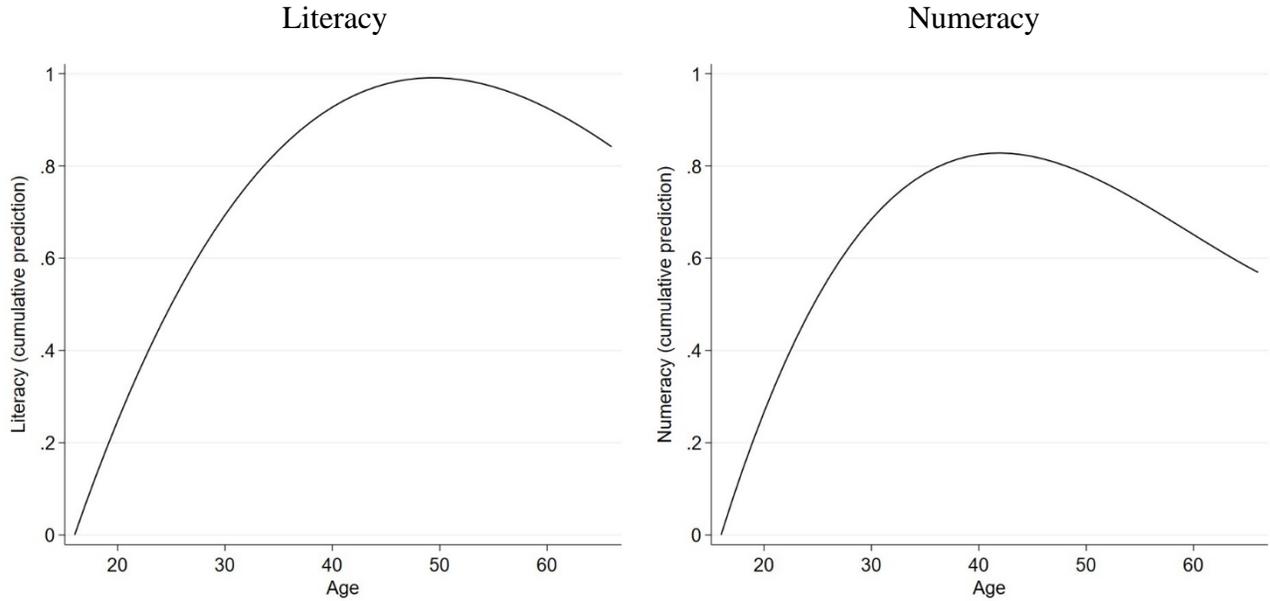

B. Marginal skill changes by age

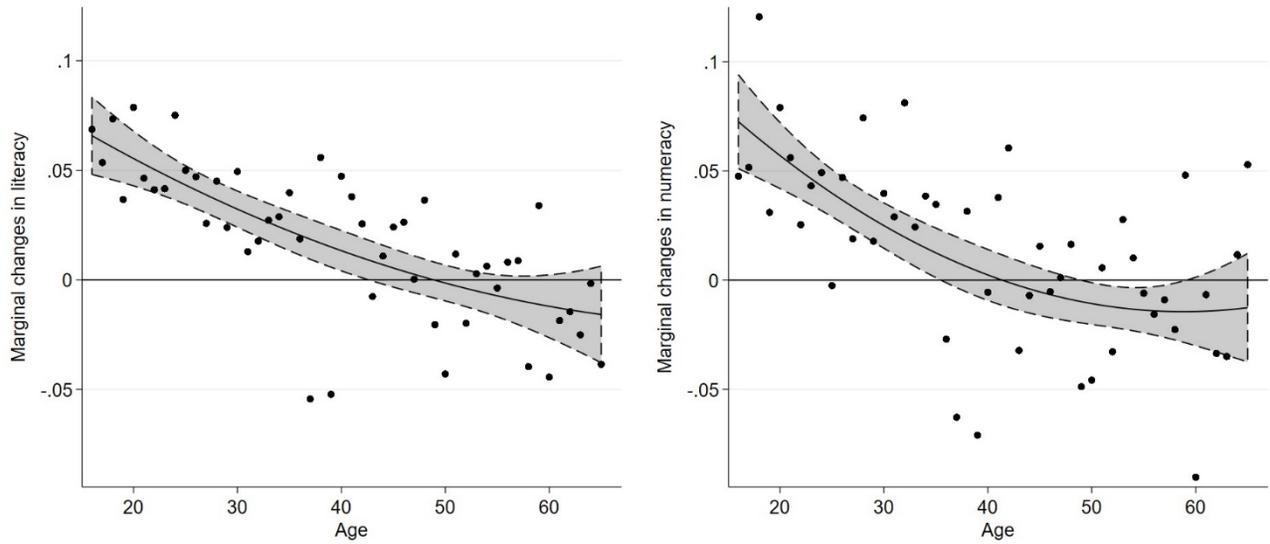

Notes: Panel A: cumulative depiction of the predicted marginal change in skills at each age. Panel B: marginal annualized change in skills between the two waves by age, adjusted for reversion to the mean. Dots: average individual marginal change in skills by age. Line: quadratic fit. Gray area: 95 percent confidence interval. Sample: employed workers, ages 16-65, weighted by sampling weights (N = 2,497). Data source: PIAAC-L.

**Figure A7: Age-skill profiles by skill usage: Sample aged 25-65**

A. Cumulative age-skill profiles

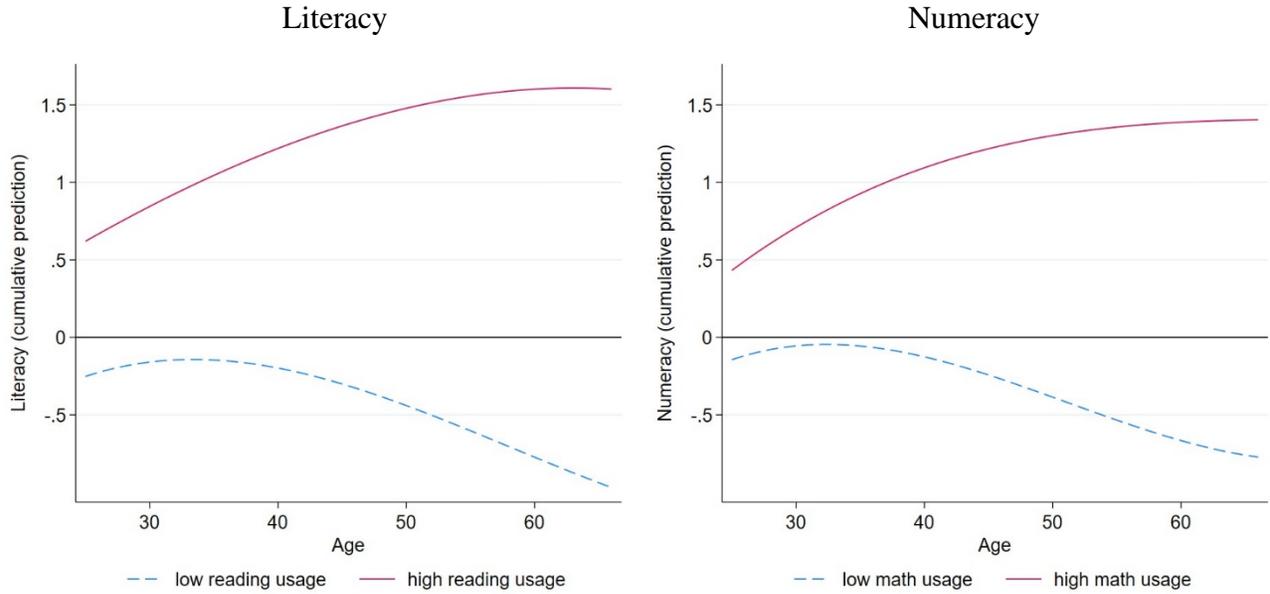

B. Marginal skill changes by age

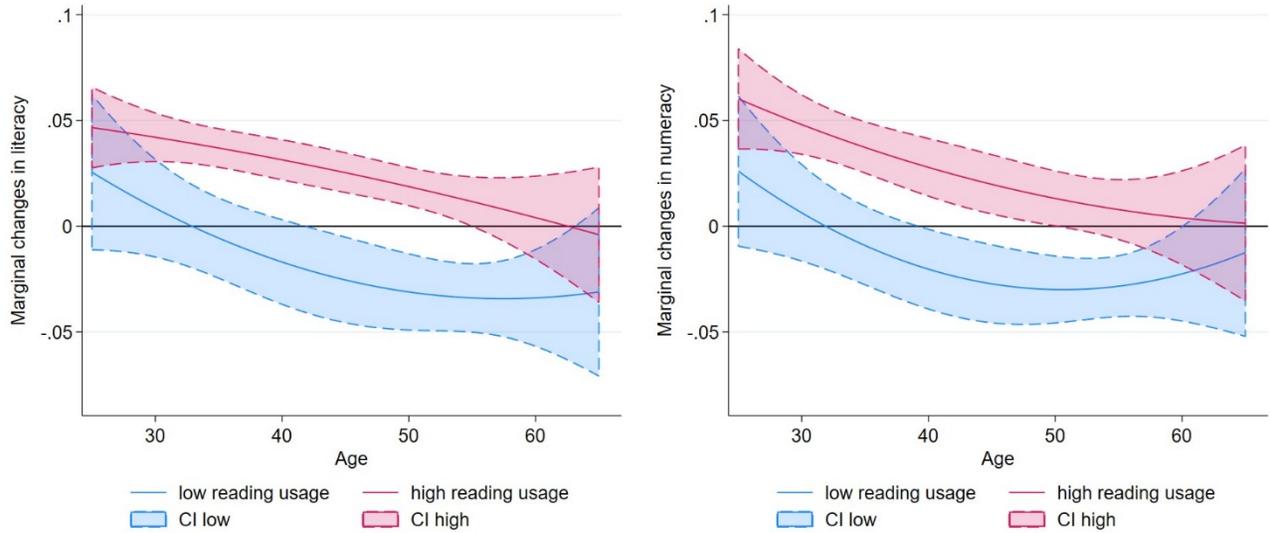

Notes: Panel A: cumulative depiction of the predicted marginal change in skills at each age. Panel B: quadratic fit (with 95 percent confidence interval) of marginal annualized change in skills between the two waves by age, adjusted for reversion to the mean. Sample split by median of skill usage at work and at home. Sample: employed workers, ages 25-65, weighted by sampling weights (N = 2,178). Data source: PIAAC-L.

**Figure A8: Age-skill profiles by skill usage at work**

A. Cumulative age-skill profiles

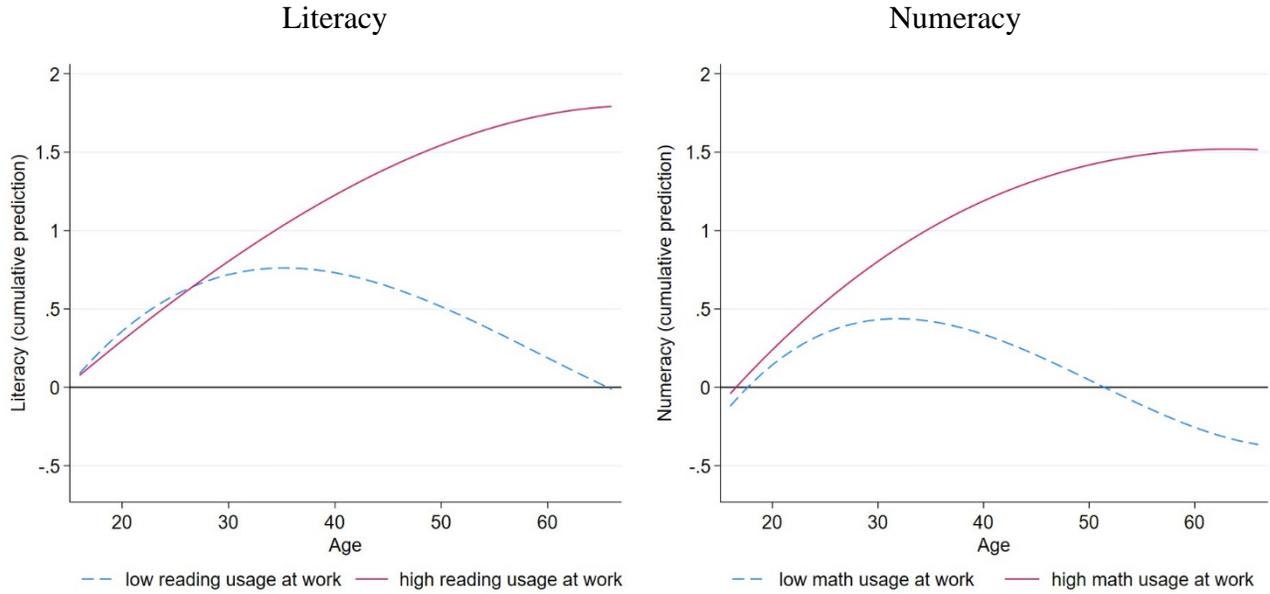

B. Marginal skill changes by age

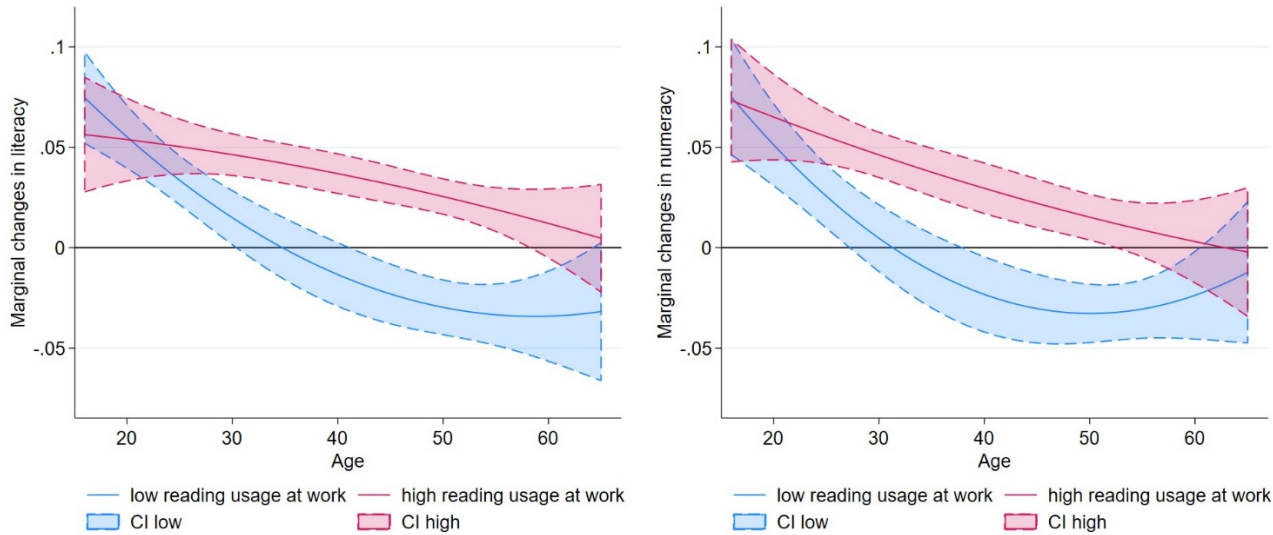

Notes: Panel A: cumulative depiction of the predicted marginal change in skills at each age. Panel B: quadratic fit (with 95 percent confidence interval) of marginal annualized change in skills between the two waves by age, adjusted for reversion to the mean. Sample split by median of skill usage at work. Sample: employed workers, ages 16-65, weighted by sampling weights (N = 2,497). Data source: PIAAC-L.

**Figure A9: Age-skill profiles by skill usage at home**

A. Cumulative age-skill profiles

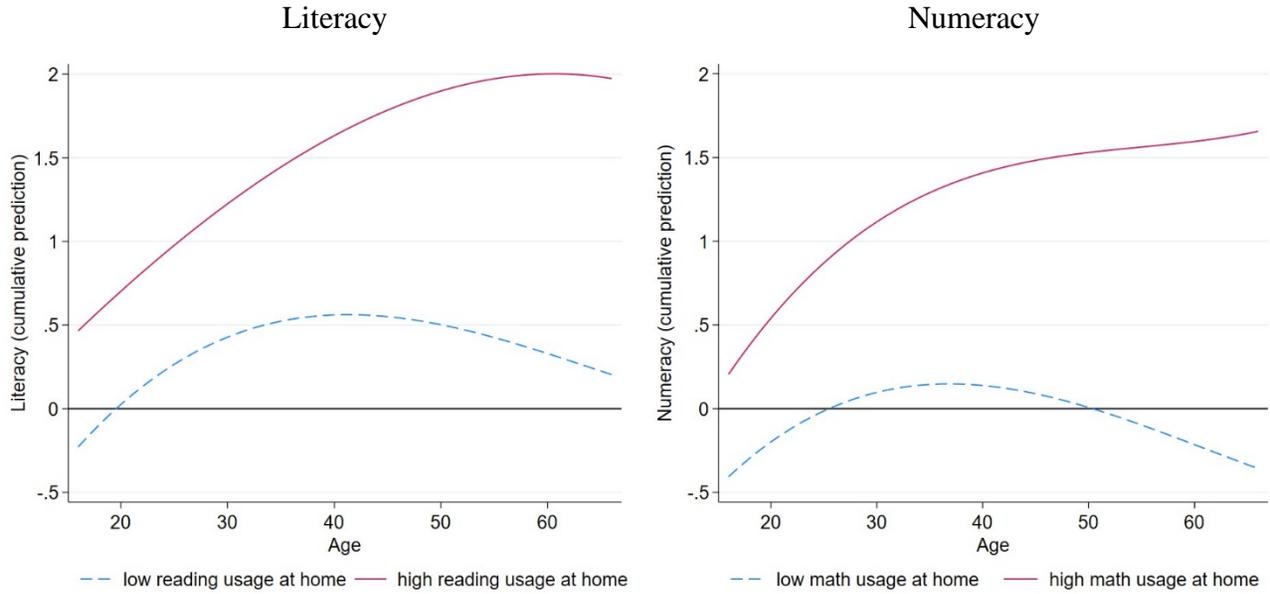

B. Marginal skill changes by age

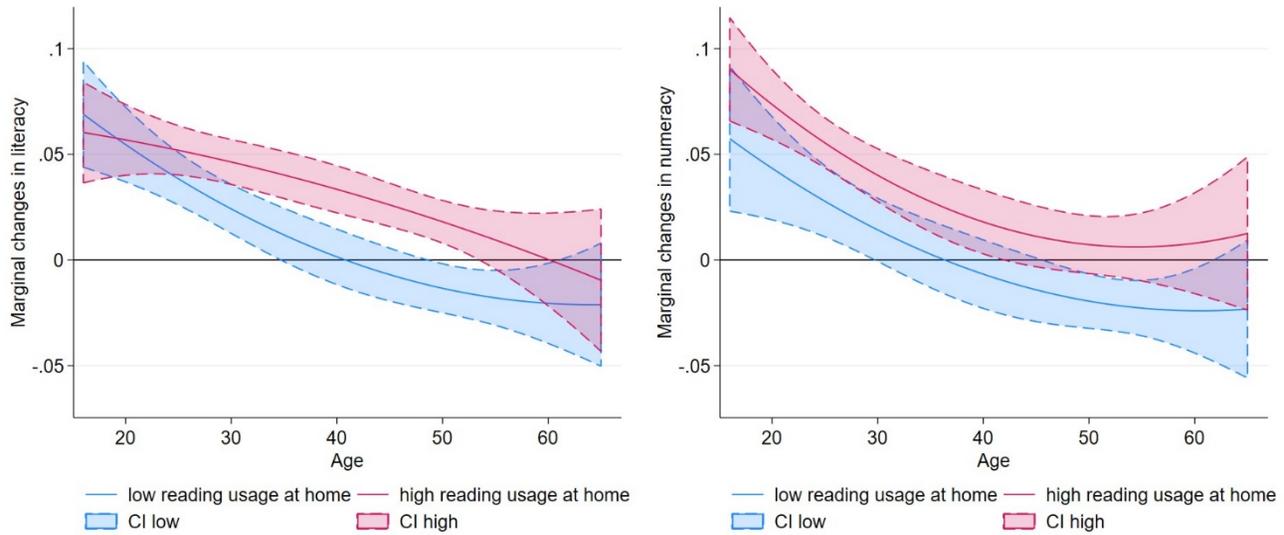

Notes: Panel A: cumulative depiction of the predicted marginal change in skills at each age. Panel B: quadratic fit (with 95 percent confidence interval) of marginal annualized change in skills between the two waves by age, adjusted for reversion to the mean. Sample split by median of skill usage at home. Sample: employed workers, ages 16-65, weighted by sampling weights (N = 2,497). Data source: PIAAC-L.

**Figure A10: Age-skill profiles by skill usage at home: Full population**

A. Cumulative age-skill profiles

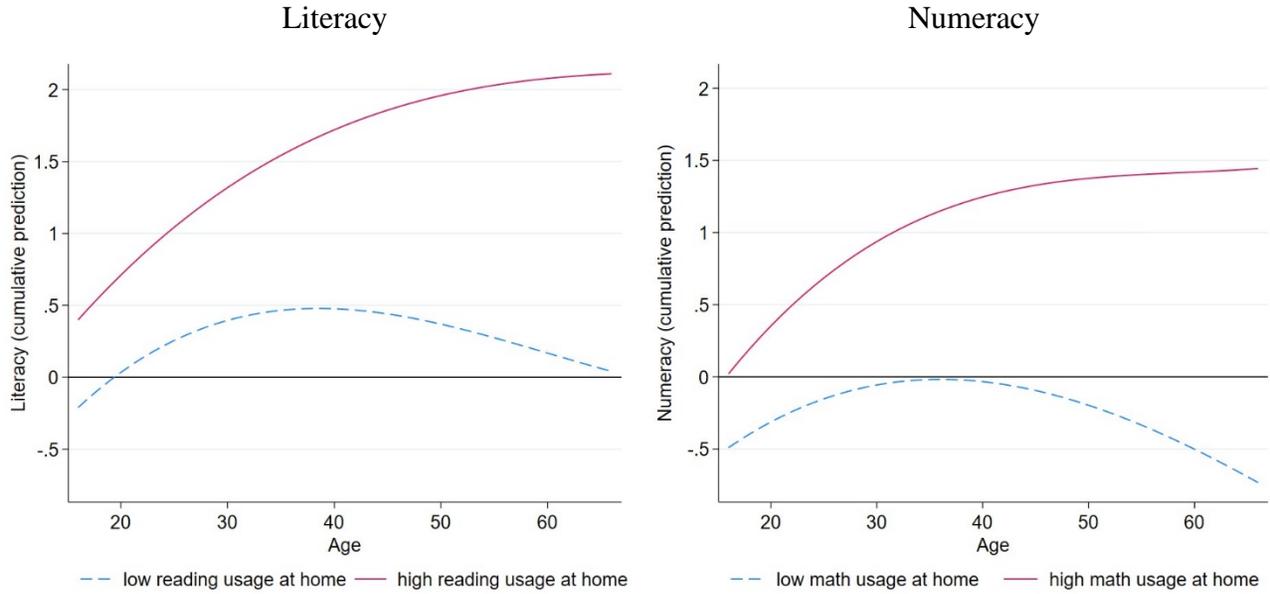

B. Marginal skill changes by age

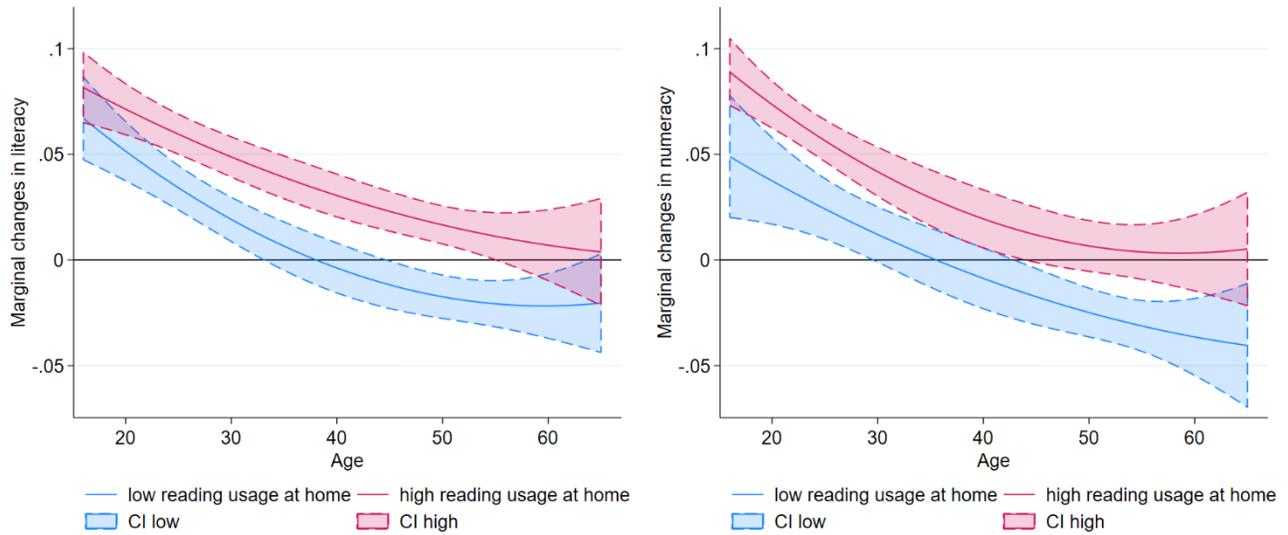

Notes: Panel A: cumulative depiction of the predicted marginal change in skills at each age. Panel B: quadratic fit (with 95 percent confidence interval) of marginal annualized change in skills between the two waves by age, adjusted for reversion to the mean. Sample split by median of skill usage at home. Sample: full population, ages 16-65, weighted by sampling weights (N = 3,263). Data source: PIAAC-L.

**Figure A11: Age-skill profiles by skill usage: Without adjustment for reversion to the mean**

A. Cumulative age-skill profiles

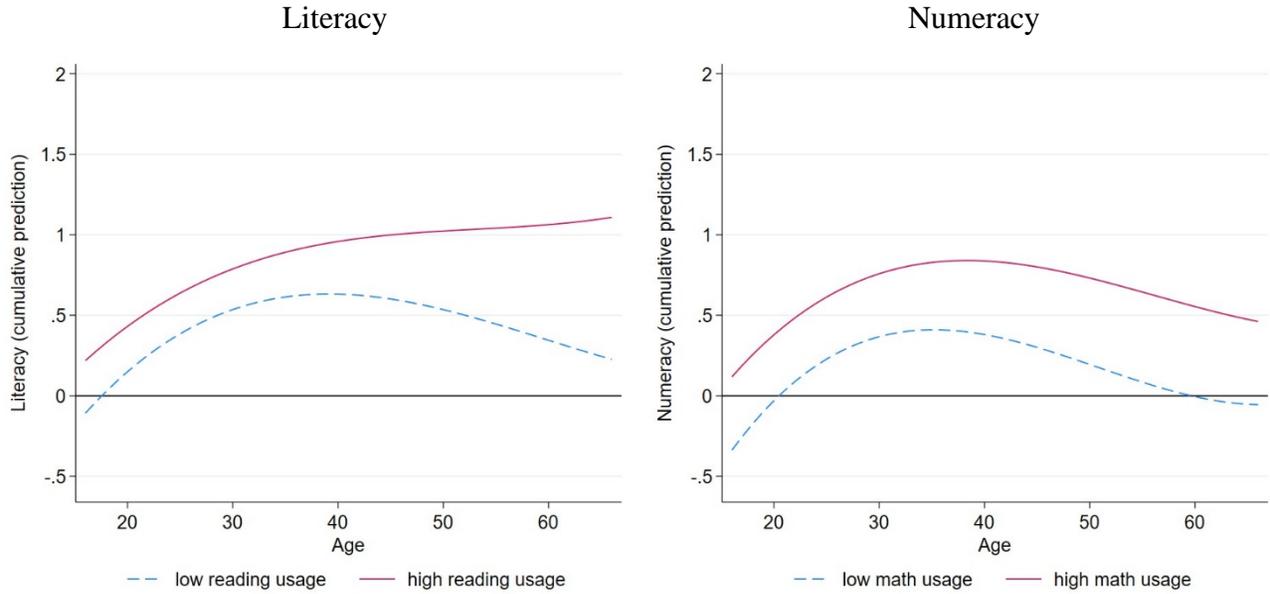

B. Marginal skill changes by age

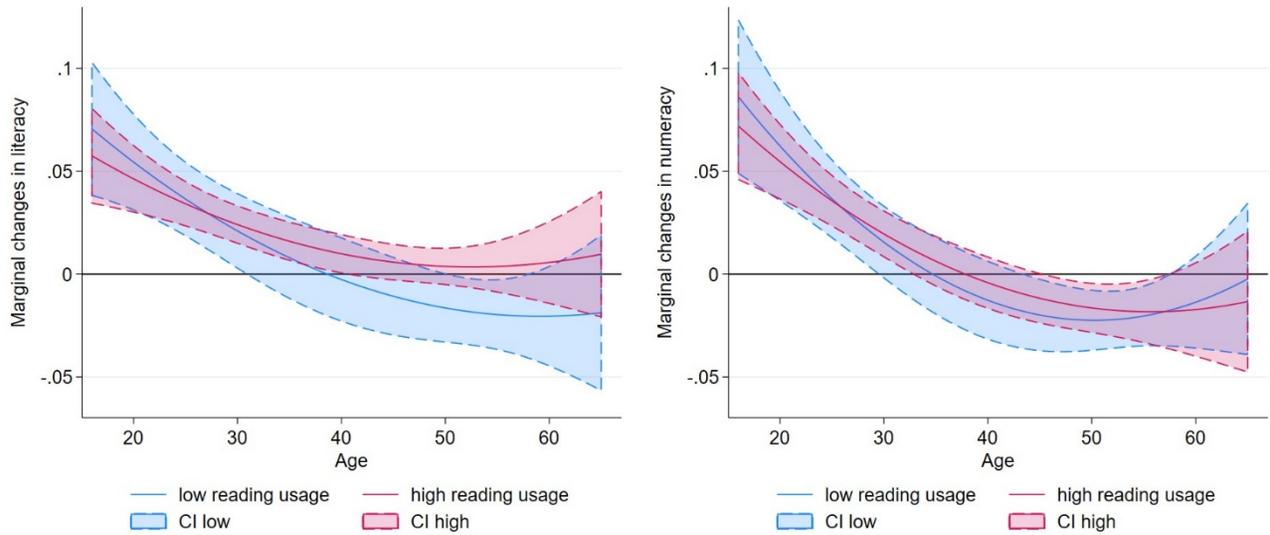

Notes: Raw scores not adjusted for reversion to the mean. Panel A: cumulative depiction of the predicted marginal change in skills at each age. Panel B: quadratic fit (with 95 percent confidence interval) of marginal annualized change in skills between the two waves by age. Sample split by median of skill usage at work and at home. Sample: employed workers, ages 16-65, weighted by sampling weights (N = 2,497). Data source: PIAAC-L.

**Figure A12: Age-skill profiles by occupation**

A. Cumulative age-skill profiles

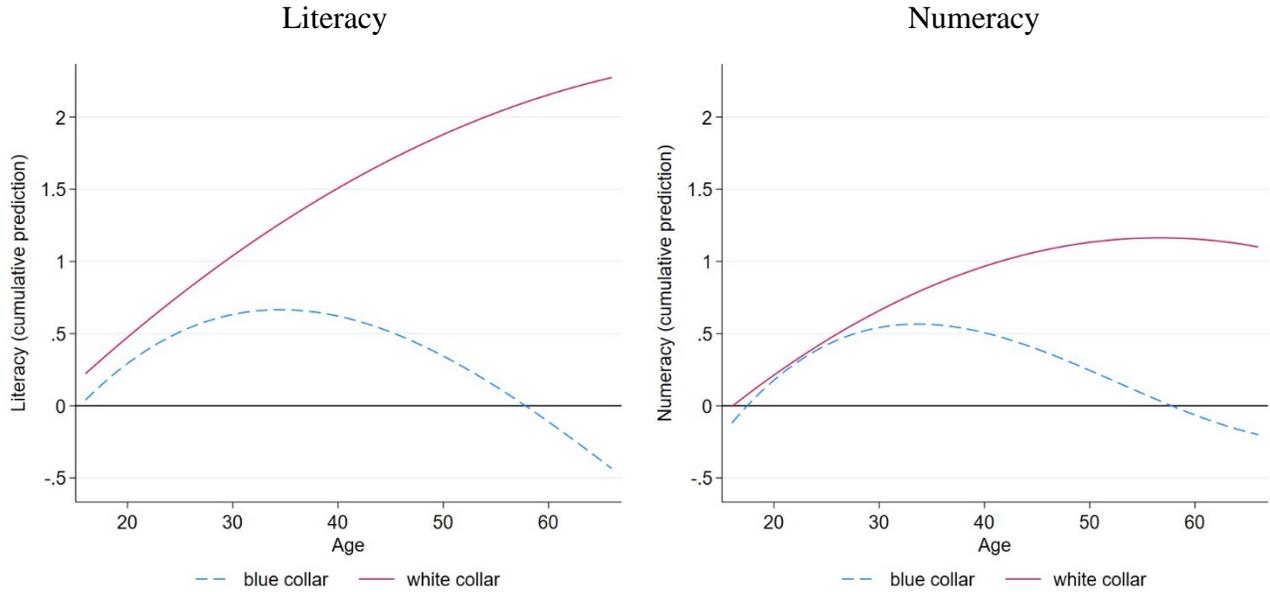

B. Marginal skill changes by age

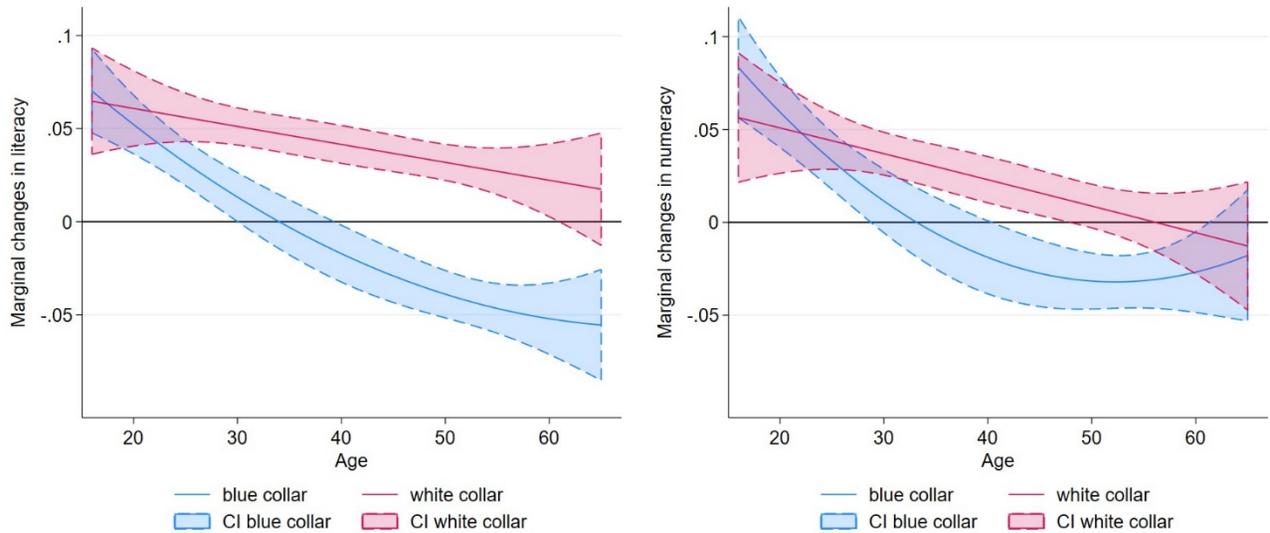

Notes: Panel A: cumulative depiction of the predicted marginal change in skills at each age. Panel B: quadratic fit (with 95 percent confidence interval) of marginal annualized change in skills between the two waves by age, adjusted for reversion to the mean. Sample split between blue- and white-collar workers. Sample: employed workers, ages 16-65, weighted by sampling weights (N = 2,497). Data source: PIAAC-L.

**Figure A13: Age-skill profiles by education**

A. Cumulative age-skill profiles

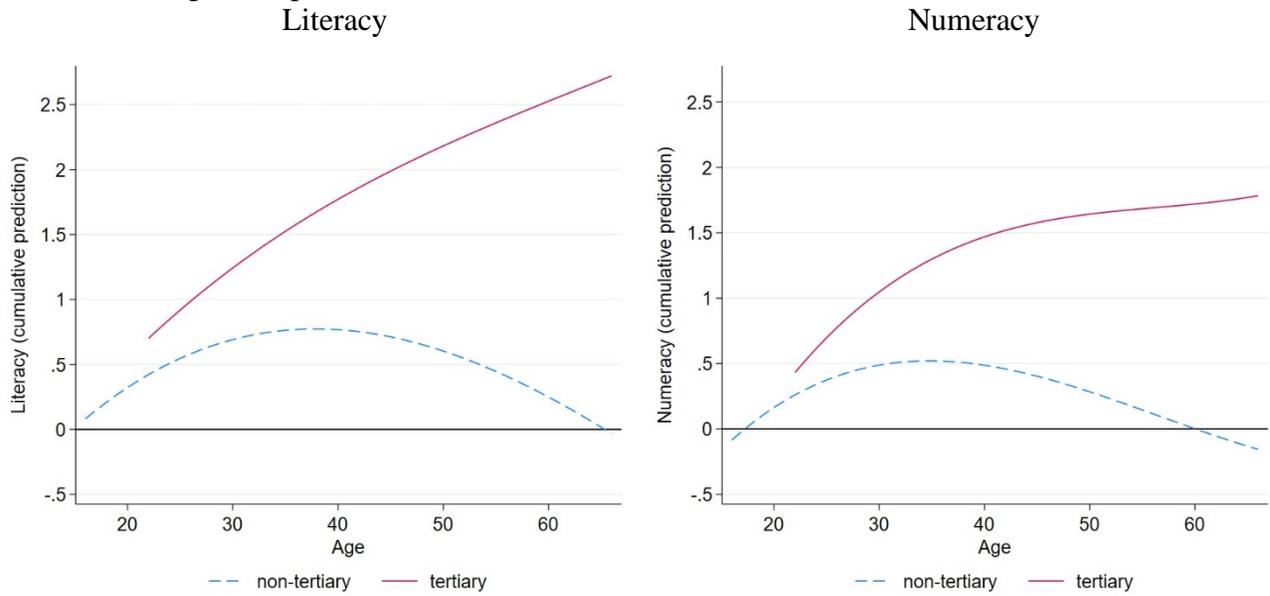

B. Marginal skill changes by age

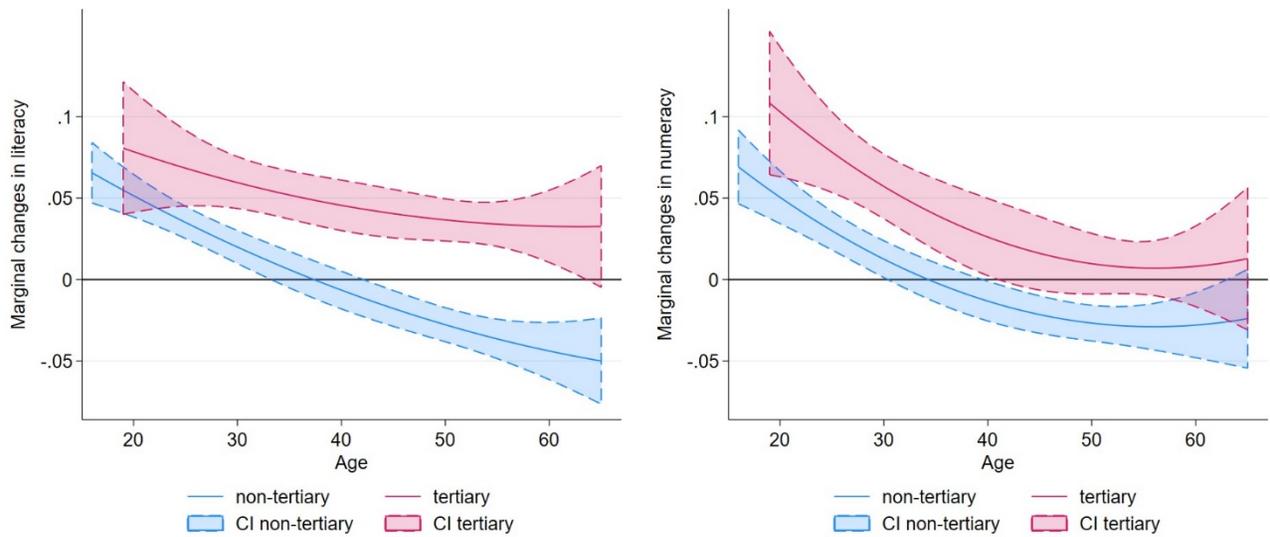

Notes: Panel A: cumulative depiction of the predicted marginal change in skills at each age. Panel B: quadratic fit (with 95 percent confidence interval) of marginal annualized change in skills between the two waves by age, adjusted for reversion to the mean. Sample split between workers with and without a tertiary education. Sample: employed workers, ages 16-65, weighted by sampling weights (N = 2,497). Data source: PIAAC-L.

**Figure A14: Age-skill profiles by gender**

A. Cumulative age-skill profiles

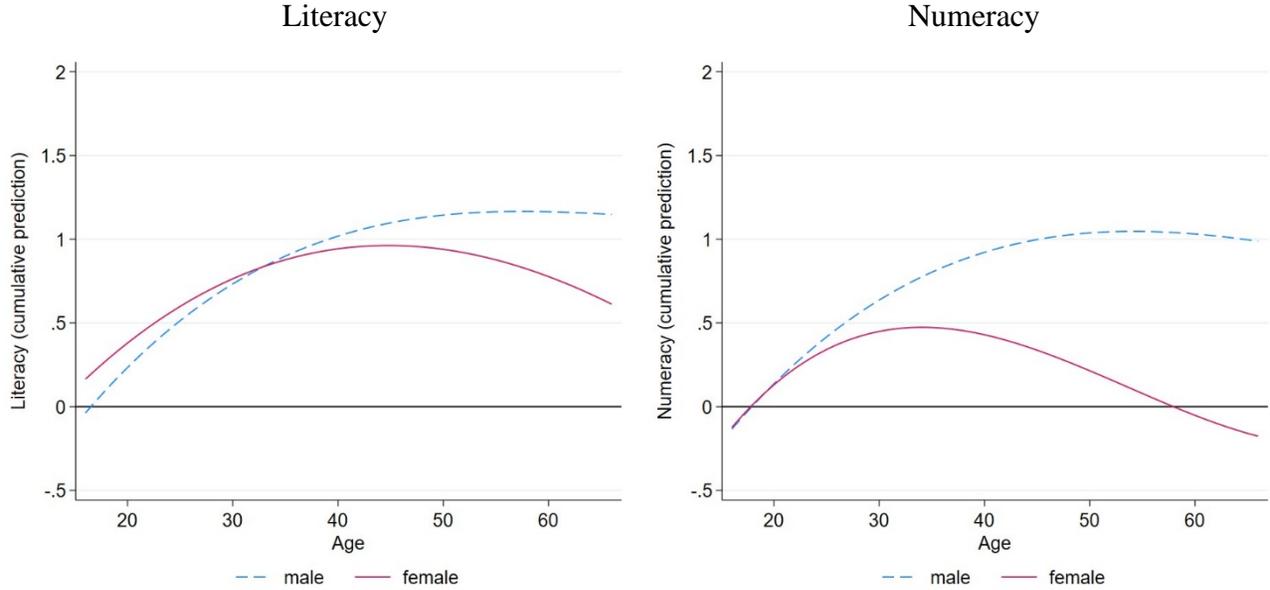

B. Marginal skill changes by age

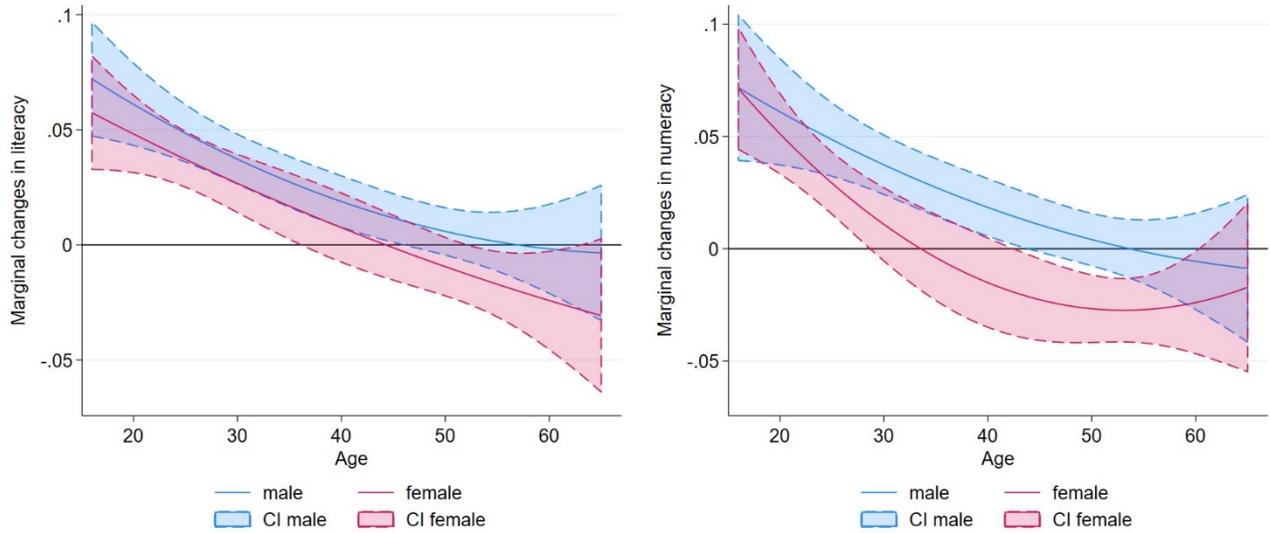

Notes: Panel A: cumulative depiction of the predicted marginal change in skills at each age. Panel B: quadratic fit (with 95 percent confidence interval) of marginal annualized change in skills between the two waves by age, adjusted for reversion to the mean. Sample split by gender. Sample: employed workers, ages 16-65, weighted by sampling weights (N = 2,497). Data source: PIAAC-L.

**Figure A15: Age-skill profiles by initial level of skills**

A. Cumulative age-skill profiles

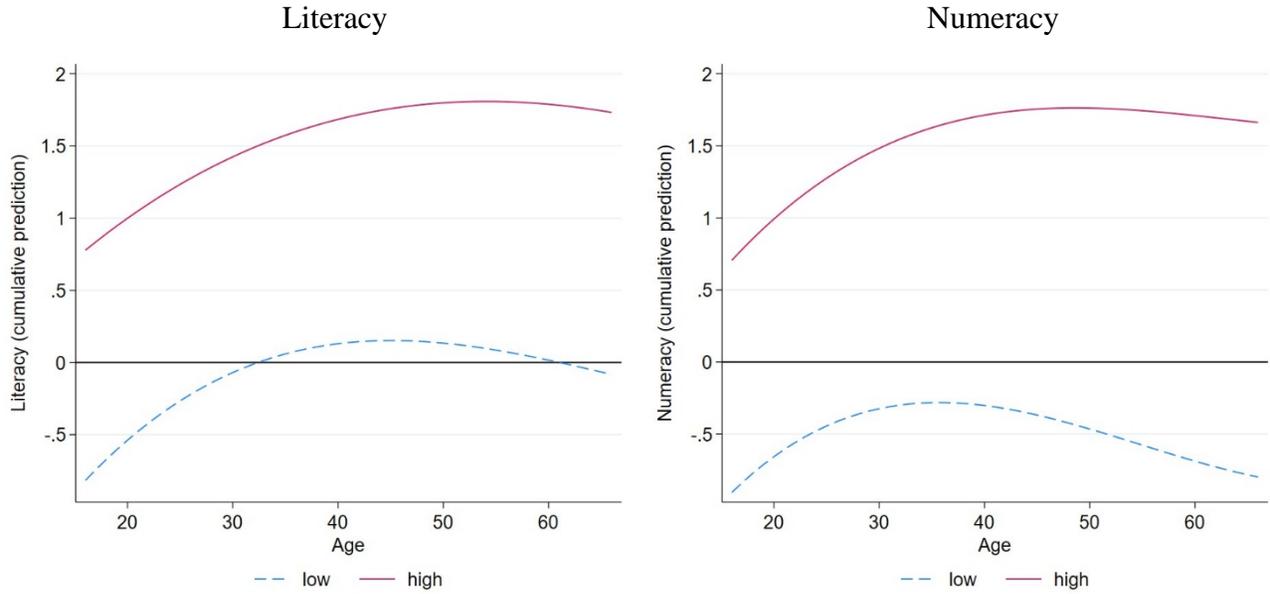

B. Marginal skill changes by age

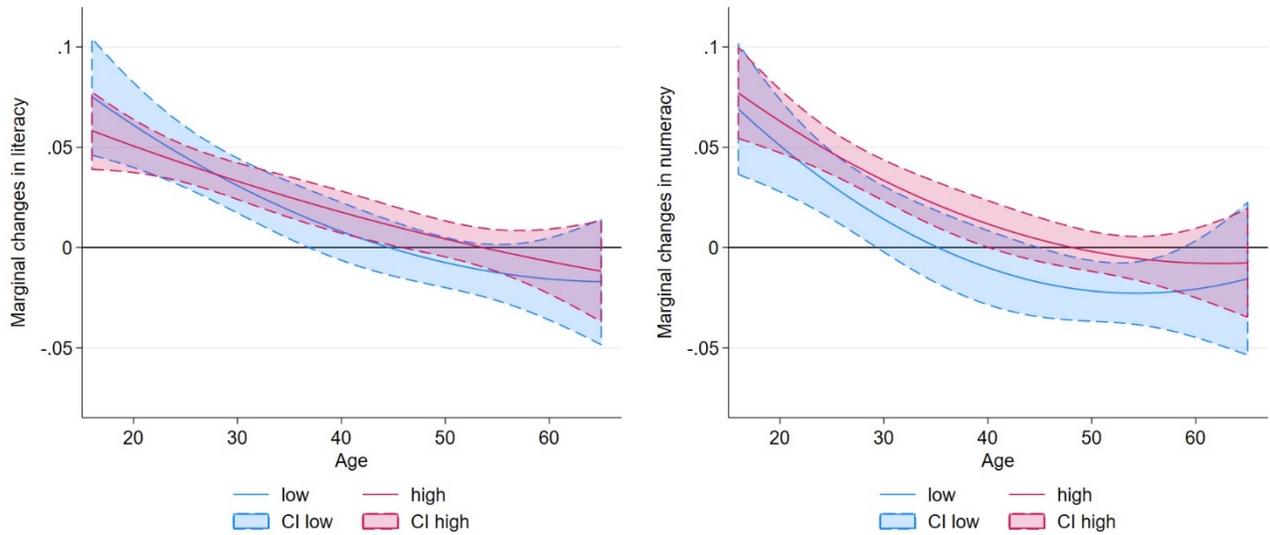

Notes: Panel A: cumulative depiction of the predicted marginal change in skills at each age. Panel B: quadratic fit (with 95 percent confidence interval) of marginal annualized change in skills between the two waves by age, adjusted for reversion to the mean. Sample split by median scores in initial (2012) wave. Sample: employed workers, ages 16-65, weighted by sampling weights (N = 2,497). Data source: PIAAC-L.

**Figure A16: Skill changes after age 40: Splits by medians of usage within subgroups**

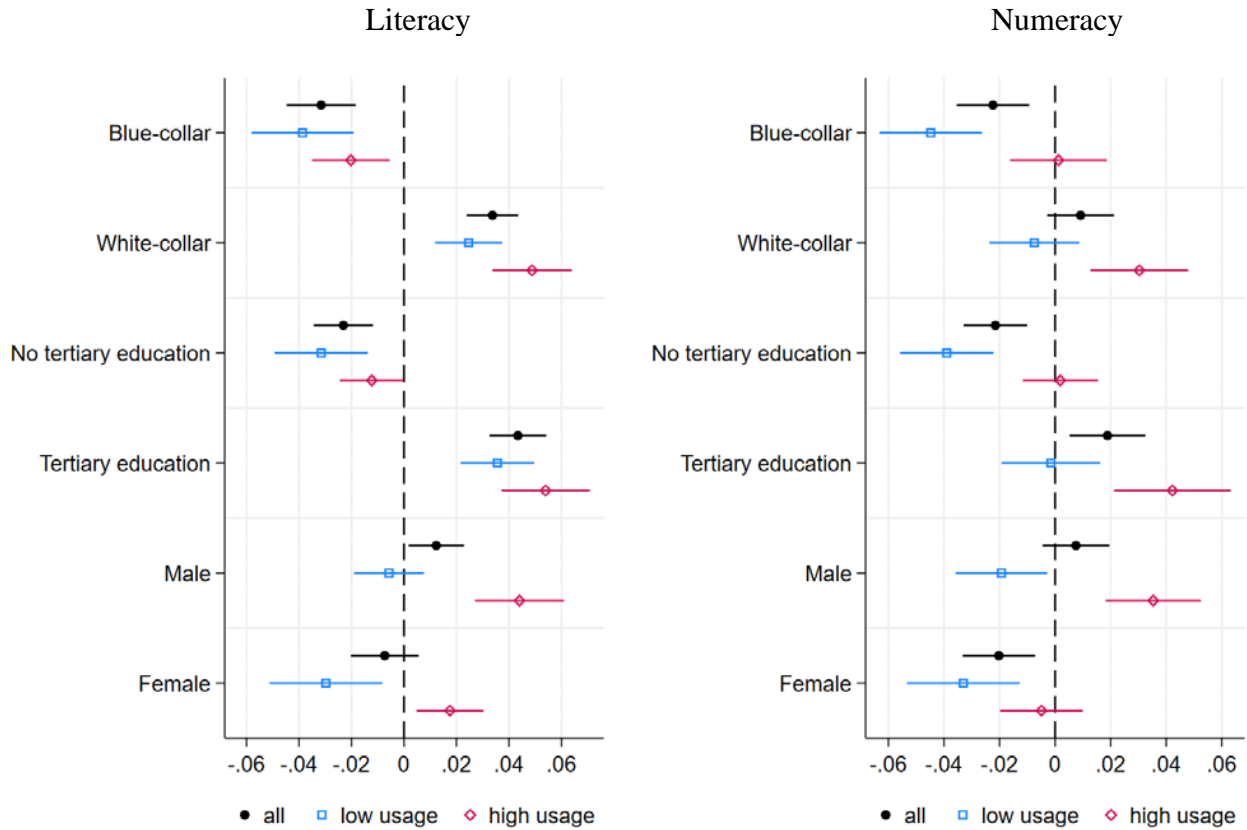

Notes: Average individual marginal annualized change in skills between the two waves, adjusted for reversion to the mean, and 95 percent confidence band. Subgroup means by blue-/white-collar occupations, (no) tertiary education, and gender, respectively. Low/high skill usage: below/above median of skill usage at work and at home within the respective subgroup. Sample: employed workers, ages 40-65, weighted by sampling weights. Data source: PIAAC-L.

**Table A1: Descriptive statistics**

|  | Full sample | | Employed sample | |
|---|---|---|---|---|
|  | Mean | Std. dev. | Mean | Std. dev. |
|  | (1) | (2) | (3) | (4) |
| Age | 41.329 | 13.625 | 41.869 | 12.259 |
| White-collar occupation | 0.378 |  | 0.501 |  |
| Tertiary education | 0.306 |  | 0.353 |  |
| Female | 0.499 |  | 0.471 |  |
| Literacy (2012) | 0 | 1 | 0.094 | 0.962 |
| Literacy (2015) | 0.049 | 1.017 | 0.126 | 0.976 |
| Change in literacy (annual) | 0.014 | 0.156 | 0.013 | 0.154 |
| Numeracy (2012) | 0 | 1 | 0.125 | 0.943 |
| Numeracy (2015) | 0.020 | 1.042 | 0.123 | 0.996 |
| Change in numeracy (annual) | 0.006 | 0.174 | 0.006 | 0.172 |
| Observations | 3,263 |  | 2,497 |  |

Notes: Variables refer to initial (2012) wave unless noted otherwise. Sample: full population and employed workers, respectively, ages 16-65, weighted by sampling weights. Data source: PIAAC-L.

**Table A2: Heterogeneity in marginal changes in skills: Different measures of skill usage**

|  | Literacy | | | | | | Numeracy | | | | | |
| --- | --- | --- | --- | --- | --- | --- | --- | --- | --- | --- | --- | --- |
|  | (1) | (2) | (3) | (4) | (5) | (6) | (7) | (8) | (9) | (10) | (11) | (12) |
| Skill usage at work | 0.079*** (0.012) |  | 0.067*** (0.013) |  |  |  | 0.073*** (0.013) |  | 0.065*** (0.012) |  |  |  |
| Skill usage at home |  | 0.071*** (0.012) | 0.036*** (0.012) |  |  |  |  | 0.058*** (0.015) | 0.031** (0.014) |  |  |  |
| Reading skill usage |  |  |  | 0.097*** (0.016) |  |  |  |  |  | 0.082*** (0.018) |  |  |
| Math skill usage |  |  |  | 0.022 (0.014) |  |  |  |  |  | 0.059*** (0.017) |  |  |
| Skill usage (full range) |  |  |  |  | 0.146*** (0.020) | 0.078*** (0.023) |  |  |  |  | 0.128*** (0.023) | 0.096*** (0.024) |
| Age and age squared | yes | yes | yes | yes | yes | yes | yes | yes | yes | yes | yes | yes |
| Background controls | no | no | no | no | no | yes | no | no | no | no | no | yes |
| $R^2$ (adj.) | 0.044 | 0.031 | 0.047 | 0.047 | 0.049 | 0.071 | 0.036 | 0.025 | 0.037 | 0.046 | 0.038 | 0.046 |
| Observations | 2,497 | 2,497 | 2,497 | 2,497 | 2,497 | 2,497 | 2,497 | 2,497 | 2,497 | 2,497 | 2,497 | 2,497 |

Notes: Least squares regressions weighted by sampling weights. Dependent variable: individual marginal annualized change in skills between the two waves, adjusted for reversion to the mean. Skill usage measures refer to average of indicators of at least monthly skill usage of battery of usage categories, except for "skill usage (full range)", which uses the underlying five-point scale of usage (ranging from never to every day), linearized from zero to one. Background controls: white-collar occupation, tertiary education, and female. Sample: employed workers, ages 16-65. Regressions use ten plausible values of skill measurement per observation (individual). Standard errors clustered at the individual level in parentheses. Significance level: *** 1 percent, ** 5 percent, * 10 percent. Data source: PIAAC-L.